\documentclass[fleqn,usenatbib,useAMS]{mnras}

\usepackage{graphicx}	
\graphicspath{{Figures/}} 
\usepackage{amsmath}	
\usepackage{amssymb}	
\usepackage{multicol}        
\usepackage{bm}		
\usepackage{pdflscape}	
\usepackage{physics}

\usepackage{siunitx}

\usepackage[T1]{fontenc}
\usepackage{ae,aecompl}

\usepackage{mathptmx}

\usepackage{color}
\definecolor{red}{RGB}{0, 0, 0}

\usepackage{cleveref}

\title[A dissipative extension to ideal hydrodynamics]{A dissipative extension to ideal hydrodynamics}

\author[M. J. Hatton and I. Hawke]{Marcus John Hatton$^{1}$\thanks{Contact e-mail: \href{hatton:mjh1n20@soton.ac.uk}{mjh1n20@soton.ac.uk}} and Ian Hawke$^{1}$
\\
$^{1}$Mathematical Sciences and STAG Research Centre, University of Southampton, Southampton, SO17 1BJ, UK}

\date{\today}
\pubyear{2023}

\begin{document}
\label{firstpage}
\pagerange{\pageref{firstpage}--\pageref{lastpage}}
\maketitle

\begin{abstract}
We present a formulation of special relativistic, dissipative hydrodynamics (SRDHD) derived from the well-established M{\"u}ller-Israel-Stewart (MIS) formalism using an expansion in deviations from ideal behaviour. By re-summing the non-ideal terms, our approach extends the Euler equations of motion for an ideal fluid through a series of additional source terms that capture the effects of bulk viscosity, shear viscosity and heat flux. For efficiency these additional terms are built from purely spatial derivatives of the primitive fluid variables. The series expansion is parametrized by the dissipation strength and timescale coefficients, and is therefore rapidly convergent near the ideal limit. We show, using numerical simulations, that our model reproduces the dissipative fluid behaviour of other formulations. As our formulation is designed to avoid the numerical stiffness issues that arise in the traditional MIS formalism for fast relaxation timescales, it is roughly an order of magnitude faster than standard methods near the ideal limit.
\end{abstract}


\section{Motivation}

Binary neutron star mergers represent complex astrophysical laboratories that probe supernuclear-density matter, strong-gravity spacetime and the origin of the heavy elements in our universe. Merging neutron stars produce multi-messenger signals comprised of extraordinary electromagnetic and gravitational wave components, as confirmed by their detection in the GW170817 merger event by the LIGO-VIRGO-Kagra collaboration~\citep{abbott_gw170817_2017}. Since then, observations by these ground-based detectors have been used to put constraints on the mass, radius and tidal deformability of neutron stars, informing us in turn about their equation of state~\citep{abbott_gw170817_2018, abbott_properties_2019}. 

To simulate these events, we require a highly non-linear, general relativistic, magneto-hydrodynamic (GRMHD) model. This is then coupled to a spacetime evolution procedure using numerical relativity, with neutrino transport and cooling schemes often added as well. For \textit{simplicity}, the fluid of the neutron star is often treated as `ideal', in that fluid stresses are purely isotropic.

Recently, however, more attention has been paid to effects resulting from non-ideal fluid behaviour, for example by~\cite{shibata_general_2017,rezzolla_physics_2018,bemfica_causality_2019,chabanov_general-relativistic_2021, pandya_conservative_2022,yang_far--equilibrium_2024}. These dissipative effects arise due to out-of-equilibrium processes which are particularly important shortly after the neutron stars merge. When next-generation, ground-based gravitational wave detectors such as Cosmic Explorer~\citep{reitze_cosmic_2019}, NEMO~\citep{ackley_neutron_2020}, LIGO-voyager~\citep{berti_snowmass2021_2022} and the Einstein Telescope~\citep{punturo_einstein_2010} come online, accurately modelling this next-to-leading-order behaviour will be essential to make precise physical inferences from observations.

Theoretical work~\citep{chugunov_shear_2005, manuel_shear_2011, schmitt_reaction_2018} and numerical investigations~\citep{hammond_thermal_2021} have been undertaken into the out-of-equilibrium state of matter and its transport properties in neutron stars. For example, Urca and reverse-Urca nuclear reactions operate at an atomic scale and may give rise to an effective bulk viscosity at the fluid scale that quantitatively affects the gravitational wave signal we obtain from the merger and its remnant's ringdown~\citep{alford_viscous_2018,most_projecting_2021,most_emergence_2022,hammond_impact_2023}. Similarly, work has been done to investigate the possible effects of both shear viscosity~\citep{duez_general_2004} and heat transport~\citep{alford_viscous_2018} in binary neutron star mergers, particularly for modulating the turbulence that ensues post-merger, both in the remnant itself, and its associated accretion disk. Viscous braking redistributes momentum in a differentially-rotating remnant, removing centrifugal support which can aid in the collapse of the core into a black hole. This produces a delayed gravitational wave emission. It is also able to provide thermal support from viscous heating, negating this effect.

One well-established model of non-ideal hydrodynamics is that of M\"uller, Israel and Stewart (MIS)~\citep{israel_nonstationary_1976,israel_transient_1979}. Its theoretical properties have received thorough investigations~\citep{molnar_numerical_2010,biswas_causality_2020,bemfica_nonlinear_2021,bemfica_first-order_2022,wagner_regime_2024} and it has been used extensively in the context of high-energy, quark-gluon-plasma (QGP) physics to model post-collision fluid evolution~\citep{del_zanna_relativistic_2013,du_31-dimensional_2020}, as well as in the astrophysical community for modelling viscous black-hole accretion~\citep{chabanov_general-relativistic_2021}, for example.

The MIS model includes viscous and heat-conductive effects in the evolved conserved and flux vectors, as well as relaxation-type sources that drive the non-ideal terms to relativistic analogues of their Navier-Stokes forms. A numerical issue arises when the dissipative relaxation timescales become small and the sources become `stiff'. The relaxation timescales tend to zero in the ideal limit, which is relevant for the majority of the lifecycle of a binary neutron star merger. One must either reduce the timestep of the simulation drastically or adopt implicit time-integrator methods to ensure accurate and stable numerical evolution. See \citet{palenzuela_beyond_2009, dionysopoulou_general-relativistic_2013, miranda-aranguren_hllc_2018, ripperda_general-relativistic_2019,wright_resistive_2020, wright_non-ideal_2020,dash_charge_2023} for examples of approaches taken to evolve stiff numerical systems. Both options increase the computational cost of simulations greatly. Sometimes, `best-of-both' implicit-explicit methods~\citep{pareschi_implicitexplicit_2005} may be used but in any case, the computational cost increases, potentially by orders of magnitude, when source terms become stiff near the ideal limit. 

This in turn limits the spatial resolution of simulations that are performed, leading to coarse numerical grids that represent fluid elements with sizes well above those that `should' be used to satisfy the fluid approximation. That is to say, there is significant variation in fluid properties occurring over lengthscales well below that of the grid cells' size. Estimates of the dissipation lengthscale above which structure can form through turbulence suggest that simulations may need to resolve scales below the cm level~\citep{radice_turbulence_2024,thompson_neutron_1993}. However the current highest-resolution simulations have fluid elements with sizes $\approx 10$m~\citep{kiuchi_global_2018}. To bridge this gap computationally is impractical for the foreseeable future.

Instead, to address this `sub-grid' behaviour, extensions to existing hydrodynamic models have recently begun being employed. These additions to the model aim to capture, at least in a statistical sense, either genuine sub-grid microphysics or mathematical artefacts resulting from the implicit filtering process introduced by coarse simulations. 

Sub-grid models are beginning to see a number of applications in modelling astrophysical systems. The general principle behind these extensions is to include additional terms into the equations of motion, aimed at capturing the effects of unresolved fluid behaviour at scales below that which can be directly resolved in a numerical simulation. The benefit of these models lies in their ability to, without greatly increased computational cost, capture the influence of unresolvable microphysics or fluctuations, at least in a statistical sense. 

A common application of sub-grid sources is in the modelling of turbulence. In large-eddy simulations, the equations of motion are explicitly redefined in terms of resolved and unresolved quantities. A closure relation is then applied that allows the sub-grid fields to be formulated in terms of the resolved ones. Using this technique, it is possible to replicate the behaviour that would result, on average, from using more fine-scale numerical grids. 

For instance,~\citet{radice_general-relativistic_2017} first applied an analogue of the classical Smagorinsky closure~\citep{smagorinsky_general_1963} to the equations of general relativistic hydrodynamics for a merger simulation, showing that by modelling the sub-grid scale turbulence, the collapse of the hyper-massive neutron star remnant is altered. Other work by \citet{vigano_grmhd_2020, carrasco_gradient_2020} uses a gradient expansion approach to prescribe the unresolved fields in the MHD equations. See~\cite{radice_turbulence_2024} for a modern review of the field.

One might ask why these sub-grid models are relevant to the non-ideal hydrodynamic formulation presented here. In~\cite{celora_covariant_2021}, it is shown that when a linear, covariant filtering operation is applied to an ideal fluid formulation, the fine-scale variation that is spatially-averaged over may be described on the coarse-scale by algebraic terms that mimic those present in a non-ideal fluid formulation. The corollary of this is that we may use our models of non-ideal hydrodynamics to describe a fluid which does not genuinely exhibit dissipative effects, at least not on the coarse scale at which we simulate it, but instead to capture unresolved effects due to resolution limitations. In effect, our sub-grid closure relation is given by the model's prescription for the non-ideal dissipation terms within it. Of course, the meaning of the `dissipative' terms changes when we do this. Instead, they now capture the effects of filtering, and follow-up work will be published investigating this.

In this paper, we develop an extension to the special relativistic, ideal hydrodynamic equations that captures the dissipative effects present in full non-ideal fluid descriptions. This extension, dubbed a dissipative extension to ideal fluid dynamics (DEIFY), is derived from first principles arguments, and as such requires no fine tuning of parameters for different astrophysical scenarios. The rest of the paper is laid out as follows. In \cref{sec:HD} we introduce the hydrodynamic models we are concerned with: first, zero'th-order ideal hydrodynamics; then, second-order, non-ideal hydrodynamics in the M{\"u}ller-Israel-Stewart 
(MIS) formulation. \Cref{sec:CE expansion} introduces the Chapman-Enskog (CE) expansion we use here and derives a number of simple models to demonstrate the pertinent points. \Cref{sec:MISCE_Derivation} presents the full `MISCE' model with its source derived from applying the CE expansion to the MIS model. In \cref{sec:results} we show results of simulations that use the MISCE formulation of dissipative hydrodynamics. In particular, we quantitatively compare results and performance with the MIS model. Our appendices cover considerations about initial data and stability for the MISCE model, as well as how one may calculate time derivatives of primitive fluid variables without using lagged updates. Finally, in \cref{sec:conclusions}, we summarise the findings of the previous sections, discuss how they fit into current astrophysical simulations, and propose the future direction of the project.

\section{Hydrodynamic Models}\label{sec:HD}
	
In this section, we outline two models of hydrodynamics that are used in relativistic astrophysics. In order to simplify the numerics in later sections and to test the validity of the method, we will limit ourselves to special relativity. In moving to a general relativistic description, only the form of the equations should change, and so the analysis we perform here should still apply. We will also adopt the Einstein summation convention over repeated indices, where Greek letters run over 4 indices (1 temporal and 3 spatial) and Roman letters run over 3 (spatial) indices. $\delta^i_j$ is the Kronecker delta (3,3)-tensor. We use units where $c = 1$ throughout.
	 
\subsection{Ideal Hydrodynamics}
\label{sec:ideal_hydro}

The first model we present is that of ideal, non-dissipative hydrodynamics. This is the simplest relativistic model of fluids that one can simulate. The stress-energy tensor for such a fluid is
\begin{equation}
    \label{eq:ideal_SET}
    T^{\mu\nu} = (\rho + p)u^\mu u^\nu + p g^{\mu\nu}
\end{equation}
where $\rho$ is the energy density of the fluid, $p$ is its pressure, $u^\mu$ is its 4-velocity and $g^{\mu\nu}$ is the metric tensor defining the spacetime geometry. The resulting equations of motion, in conservative form, are 
\begin{equation}
      \label{eq:Euler_conslaw}
      \partial_t \begin{pmatrix}
      D \\ S_j \\ \tau \end{pmatrix} + \partial_i \begin{pmatrix}
      D v^i \\ S_j v^i + p \delta^i_j \\ \tau v^i + pv^i
      \end{pmatrix} = 0
\end{equation}
where the three conserved quantities, \{$D$, $S_j$, $\tau$\}, correspond to the fluid density, specific momentum in the $j^{th}$-direction and kinetic energy density respectively and are related to the primitive quantities, \{$n, v_j, p$\}, namely the baryon number density, fluid 3-velocity and hydrodynamic pressure via
\begin{subequations}
  \begin{align}
    D &= n W, \\
    S_j &= (\rho + p)W^2 v_j, \\
    \tau &= (\rho + p)W^2 - D.
  \end{align}
\end{subequations}
Additionally, we have that $\rho = m n (1 + \epsilon)$ where $m$ is the mass per baryon, and $\epsilon$ the specific internal energy. The hydrodynamic pressure $p$ is given by an equation of state to close the system: numerically we often write $p \equiv p(n, \epsilon)$ but here it is more convenient to use $p \equiv p(n, \rho)$. Specifically, we will use a Gamma-law equation of state throughout this paper of the form $p = (\Gamma - 1) (\rho - m n)$.

Two more important thermodynamic quantities of interest are specific enthalpy $h = 1 + \epsilon + p / (mn) = (\rho + p) / (m n)$ and temperature $T = p / n$ (as well as its inverse $\beta = m / T$). Note that we assume a uniform baryon mass so scale it out of the equations such that $m = 1$ throughout this paper. Finally, the spatial three-velocity $v_j$ and the Lorentz factor $W = (1 - v_j v^j)^{-1/2}$ make up the four-velocity $u_\mu = W (1, v_j)$.
	
\subsection{Dissipative Hydrodynamics Within M{\"u}ller-Israel-Stewart Formalism}

The second model we present is that describing a non-ideal fluid with a stress-energy tensor given by
\begin{equation}
    \label{eq:non-ideal_SET}
    T^{\mu\nu} = (\rho + p + \Pi)u^\mu u^\nu + (p + \Pi) g^{\mu\nu} + q^\mu u^\nu + q^\nu u^\mu + \pi^{\mu\nu},
\end{equation}
where the new dissipative terms are the bulk viscosity pressure $\Pi$, the heat flux vector $q^\mu$, and the shear viscosity tensor $\pi^{\mu\nu}$. The first of these encapsulates isotropic stresses (compression and expansion). The second, momentum transport orthogonal to fluid's velocity. And the third, anisotropic stresses within the fluid. To set their form, and derive the equations of motion of the fluid, we follow the M{\"u}ller-Israel-Stewart (MIS)~\citep{israel_nonstationary_1976,israel_transient_1979} formalism. Their approach, in words, involves first performing a gradient expansion of entropy-generating terms to second-order. Then, applying the second law of thermodynamics such that the entropy-generation rate is always non-negative and using this condition to set the form of the non-ideal, dissipative terms.  Finally, the additional degrees of freedom this introduces are wrapped-up into six non-ideal coefficients (three dissipation `strengths' and three timescales). This gives us the model that is our starting point and the one to which we will apply a Chapman-Enskog (CE) type expansion, eventually giving us our new `MISCE' model.

The entire MIS equations, in balance-law form, are given as
\begin{equation}
    \partial_t \bm{G} + \partial_i \bm{F}^{(i)} = \bm{S}
\end{equation}
where
\begin{subequations}\label{eq:IS Full}
  \begin{align}
      \bm{G} &= \begin{pmatrix}
      D \\ S_j \\ \tau \\ U \\ Y_j \\ Z_{jk}
      \end{pmatrix} = \begin{pmatrix}
      n W \\ (\rho + p + \Pi) W^2 v_j + W (q_0 v_j + q_j) + \pi_{0j} \\ (\rho + p + \Pi) W^2 + 2 q_0 W - (p + \Pi - \pi_{00}) - D \\ n W \Pi \\ n W q_j \\ n W \pi_{jk}
      \end{pmatrix}, \\
      \bm{F}^{(i)} &= \begin{pmatrix}
      D v^i \\ S^i_j \\ S^i - D v^i \\ U v^i \\ Y_j v^i \\ Z_{jk} v^i
      \end{pmatrix}, \\
      \bm{S} &= \begin{pmatrix}
      0 \\ 0 \\ 0 \\ \tfrac{n}{\tau_{\Pi}} ( \Pi_{\text{NS}} - \Pi ) \\ \tfrac{n}{\tau_{q}} ( q_{j, \text{NS}} - q_j ) \\ \tfrac{n}{\tau_{\pi}} ( \pi_{jk, \text{NS}} - \pi_{jk} ),
      \end{pmatrix}
  \end{align}
\end{subequations}
and 
\begin{equation}
    S^i_j = (\rho + p + \Pi) W^2 v^i v_j + W(q^i v_j + q_j v^i) + (p + \Pi) \delta^i_j + \pi^i_j. 
\end{equation}
The heat flux and shear viscosity are orthogonal to the four velocity on all indices, and the shear viscosity is trace free, implying
\begin{subequations}
  \begin{align}
      q_0 &= v^k q_k, \\
      \pi_{0j} &= v^k \pi_{kj}, \\
      \pi_{j0} &= v^k \pi_{jk}, \\
      \pi_{00} &= -\pi^k_k.
  \end{align}
\end{subequations}
The first-order, relativistic ``Navier-Stokes'' terms to which the dissipative system relaxes are  
\begin{subequations}
  \begin{align}
      \Pi_{\text{NS}} &= - \zeta \Theta, \\
      q_{j, \text{NS}} &= - \kappa T (\partial_j \log T + a_j), \\
      \pi_{jk, \text{NS}} &= -2 \eta \sigma_{jk},
  \end{align}
\end{subequations}
where the non-ideal coefficients of bulk viscosity, heat conductivity and shear viscosity are $\zeta, \kappa$ and $\eta$, respectively, which we may collectively represent as ${\xi}$. The following quantities,
\begin{subequations}
  \begin{align}
      \Theta &= \partial_\mu u^\mu, \\
      a_\mu &= u^\nu \partial_\nu u_\mu, \\
      \sigma_{\mu\nu} &= \left( \partial_\mu u_\nu + \partial_\nu u_\mu - \tfrac{2}{3} \eta_{\mu\nu} \Theta \right)
  \end{align}
\end{subequations}
are the expansion, acceleration and shear of the 4-velocity. 

The first three conserved quantities (\{$D$, $S_j$, $\tau$\}), and their associated equations of motion, form the non-stiff subsystem, labeled $\bm{q}$, which reduces to the Euler equations  (in the form of~\cref{eq:Euler_conslaw}) in the ideal limit of zero dissipation ($\xi \to 0$). The remaining three conserved quantities, \{$U$, $Y_j$, $Z_{jk}$\} are labeled $\bar{\bm{q}}$. These quantities are evolved with a source that is proportional to the reciprocal of a possibly small timescale and may therefore represent a stiff subsystem. Note that the terms $\bm{q}, \bar{\bm{q}}$ should not be confused with the heat flux, which will intentionally be written in component form, $q_j$, for clarity. 

\subsubsection*{Asymptotic Behaviour}

One major difference between the two models of hydrodynamics that are presented here lies in the form of their source terms. Whilst the ideal model's source vector is entirely zero, the Israel-Stewart model's source vector has non-zero sources for the dissipative evolution components. These terms are proportional to the reciprocal of the relaxation timescales, $\tau$. In the limit as $\tau \to 0$, any deviation of the dissipative variables from their equilibrium Navier-Stokes form will be instantaneously quenched. This represents a reduction to a first-order theory, essentially a relativistic version of the classical, dissipative Navier-Stokes equations. However, we may link the dissipation timescales and  strengths, for example through thermodynamic relations for a Boltzmann gas as in~\cite{israel_nonstationary_1976} where 
\begin{subequations}
    \label{eq:betas}
    \begin{align}
        \tau_{\Pi} &= \zeta \beta_0, \\ \tau_q &= \kappa \beta_1 T, \\ \tau_{\pi} &= 2 \eta \beta_2
    \end{align}
\end{subequations}
and the $\beta$ terms are non-negative thermodynamic functions of the enthalpy, temperature and pressure given therein. Alternatively, there are analytic bounds on the ratio of dissipation strengths:timescales from enforcing causality due to recent work by~\cite{heller_rigorous_2023}. One can see this practically from the fact that dissipation modifies the characteristic propagation speeds of waves travelling in the fluid (sound speeds). For instance, by a factor $\propto \sqrt{\zeta/\tau_{\Pi}}$ for bulk viscosity as seen in~\cite{chabanov_general-relativistic_2021}. This, of course leads to a divergence when we take the $\tau \to 0$ limit without taking the $\xi \to 0$ limit along with it. In reality of course, there are no instantaneous processes. However, there are physically- and mathematically-motivated reasons why taking the instantaneous-relaxation limit also implies taking the zero-dissipation limit, in which case one recovers `zeroth-order' (ideal) hydrodynamics as in~\cref{sec:ideal_hydro}. 

\subsubsection*{Numerical difficulties}

When the timescales that the source acts on are shorter than the timestep of the simulation, $\tau \lesssim \Delta t$, the system is said to be \textit{stiff}. In order to maintain a stable evolution, one may either reduce the size of the timestep used in the simulation, or employ a set of implicit or semi-implicit time integrators such as those seen in~\cite{pareschi_implicitexplicit_2005}. In the first case where the timestep, the execution time will increase by a factor $\approx \Delta t / \tau$, making it impractical for dissipation acting well below the hydrodynamic timescale. In the latter case where (semi-) implicit methods are used, the time-integrator algorithm is considerably, maybe orders of magnitude, more costly.

Hence, our motivation is to find a source term that captures dissipative behaviour but avoids the numerical difficulties of the stiff MIS system of equations. The following section will derive such a source term using a Chapman-Enskog-type analysis. 

First we introduce the notation that will be used. We can re-write the conservative form of the MIS equations, \cref{eq:IS Full} in the following, more compact way:
\begin{subequations}  \label{eq:cons law}
    \begin{align}
        \partial_t \bm{q} + \partial_i \bm{f^i}(\bm{q}, \overline{\bm{q}})&=\bm{s}(\bm{q}, \overline{\bm{q}}) = \bm{0} \label{eq:cons law q}, \\
        \partial_t \overline{\bm{q}} + \partial_i \overline{\bm{f}}^i(\bm{q}, \overline{\bm{q}})&= \frac{\overline{\bm{s}}(\bm{q}, \overline{\bm{q}})}{\tau}, \label{eq:cons law qbar}
    \end{align}
\end{subequations}
where we indicate equations which become stiff as $\tau \to 0$ with an over-bar. This means that $\overline{\bm{q}} = \{U, Y_j, Z_{jk}\}$ with the corresponding fluxes, $\overline{\bm{f}^i}(\bm{q}, \overline{\bm{q}})$, and sources, $\overline{\bm{s}}(\bm{q}, \overline{\bm{q}})$, taken from \cref{eq:IS Full}. The remaining conserved variables are non-stiff in the ideal limit, and denoted $\bm{q} = \{D, S_j, \tau\}$. We will also denote the vector of primitive variables present in ideal hydrodynamics as $\bm{w} = \{p, \rho, n, v_x, v_y, v_z\}$ and the dissipative primitive variables as $\overline{\bm{w}} = \{q_j, \Pi, \pi_{jk}\}$.

\section{Chapman-Enskog Expansion}\label{sec:CE expansion}
	
In this section we will use a type of Chapman-Enskog (CE) method of expansion to derive the form of the DEIFY source term. A relevant application of this type of expansion is presented in~\citet{leveque_finite_2002}, in which LeVeque demonstrates how a coupled system of balance law equations may be reduced to a single, modified system. One of the coupled equations contains a potentially-stiff source term, whilst the other is an advection equation. The reduced system is a balance law with a derivative source term that is non-stiff. Hence, this example represents a simplified version of the system we have here, seen in~\cref{eq:cons law}. Next, we will show this how this method works for a heat-flux model obtained from the MIS equations.

\subsection{A Simple Heat Model}

To demonstrate the approach, we first apply it to a simple model governing two variables: the temperature, $T$, and the spatial heat flux, $q_i$. To obtain this model, which will be familiar once derived, we apply a number of  simplifying assumptions to  the MIS model.

We work with a static fluid such that the 3-velocity and, hence, the bulk and shear viscosity all vanish. All Lorentz factors become unity and the orthogonality relation $v_\mu q^\mu = 0$ means that $q_0 = 0$. We treat our particle number current, $n$, and hydrostatic pressure, $p$, as constants and hence the density, $\rho$, is now purely a function of the temperature i.e.\ $\rho \equiv \rho(T)$ and may be scaled out of the equations. After setting any remaining constant terms to one, we arrive at
\begin{subequations}
    \begin{align}
        \partial_t T + \partial_i q^i &= 0, \label{eq:toy_heat_a} \\
        \partial_t q_j &= \frac{1}{\tau_q} (q_{j, \text{NS}} - q_j).\label{eq:toy_heat_b}
    \end{align}
\end{subequations}
We note that the acceleration term usually present in the heat-flux's source will vanish so that $q_{j, \text{NS}} \to -\kappa \partial_j T$ and we obtain a rather simple pair of equations where the first has no source and the second, the `Maxwell-Cattaneo' equation~\citep{cattaneo_sulla_1948}, has no flux:
\begin{subequations}
    \label{eq:toy_heat}
    \begin{align}
        \partial_t T + \partial_i q^i &= 0, \\
        \partial_t q_j &= -\frac{1}{\tau_q} (\kappa \partial_j T + q_j).
    \end{align}
\end{subequations}

When $\tau_q$ is small, the heat flux, $q_j$, will relax to its equilibrium value, $q_{j, \text{NS}}$, rapidly, with small deviations being modulated by the size of $\tau_q$. Hence, we first write the non-ideal variable $q_j$ that we wish to eliminate from the system in terms of its equilibrium value $q_{j, \text{NS}} = -\kappa \partial_j T$ and a (small) correction term of order $\tau_q$, so
\begin{equation}
    \label{eq:toy_heat_CE_1}
    q_j = q_{j, \text{NS}} + \tau_q q^{(1)}_j,
\end{equation}
where $q^{(1)}_j$ is to be determined. We can then write the pair of equations, to first order in $\tau_q$, as
\begin{subequations}
  \label{eq:toy_heat_CE_2}
  \begin{align}
      \label{eq:toy_heat_CE_2a}
      \partial_t T + \partial_i q^i_{\text{NS}} &= - \partial^i \left( \tau_q q^{(1)}_i \right), \\
      \label{eq:toy_heat_CE_2b}
      \partial_t q_{j, \text{NS}} &= -q^{(1)}_j.
  \end{align}
\end{subequations}
By using the explicit form for the equilibrium value $q_{j, \text{NS}} = -\kappa \partial_j T$ we can write this as
\begin{subequations}
  \label{eq:toy_heat_CE_3}
  \begin{align}
      \label{eq:toy_heat_CE_3a}
      \partial_t T &= \partial^i \left( \kappa \partial_i T \right) - \partial^i \left( \tau_q q^{(1)}_i \right), \\
      \label{eq:toy_heat_CE_3b}
      \partial_t \left( -\kappa \partial_j T \right) &= -q^{(1)}_j
  \end{align}
\end{subequations}
to obtain an expression for $q^{(1)}_j$, but one which includes a temporal derivative.

For simplicity we will assume that $\kappa$ and $\tau_q$ are constants in time and space. By commuting the temporal and spatial derivatives in~\cref{eq:toy_heat_CE_3b}, we can now substitute the leading order form (zero'th order in $\tau_q$) of the equation of motion for $T$, \cref{eq:toy_heat_CE_3a}, into the relaxation equation~\cref{eq:toy_heat_CE_3b}, to determine the correction $q^{(1)}_j = \kappa^2 \partial_j^{(3)} T$ as purely spatial derivatives. Inserting this result back into~\cref{eq:toy_heat_CE_3a}, and writing the result in one spatial dimension, we finally have the CE form
\begin{equation}
    \label{eq:toy_heat_CE_simple}
    \partial_t T = \kappa \partial^{(2)}_x T - \kappa^2 \tau_q \partial^{(4)}_x T + \mathcal{O}(\tau_q^2). 
\end{equation}
Note that this result is an evolution equation written purely in terms of the temperature, $T$, and is half the size of the original system given by~\cref{eq:toy_heat}. This reduced model is non-stiff (as the source term is multiplied by the small timescale $\tau_q$, not by its reciprocal), but may lead to other numerical problems due to the higher derivatives. These features will hold true when we apply the CE expansion method to the full MIS equations, whereby dissipation will be modelled using only the primitive variables and their derivatives, and the system size will be reduced significantly.

\Cref{eq:toy_heat_CE_simple} is essentially a power series expansion in $\{ \kappa, \tau\}$ where the leading term alone ($\mathcal{O}(\kappa)$) gives us the 1D heat equation. With the next-to-leading order term, ($\mathcal{O}(\tau_q \kappa^2)$) we have a linear, diffusion-retention equation. That is to say, the second-order derivatives represent diffusive effects that spread heat isotropically and the fourth-order derivatives retain heat locally. The numerical significance of these higher-order terms are described well in~\citep{bevilacqua_significance_2011}.

\begin{figure}
    \centering
    \includegraphics[scale=0.6]{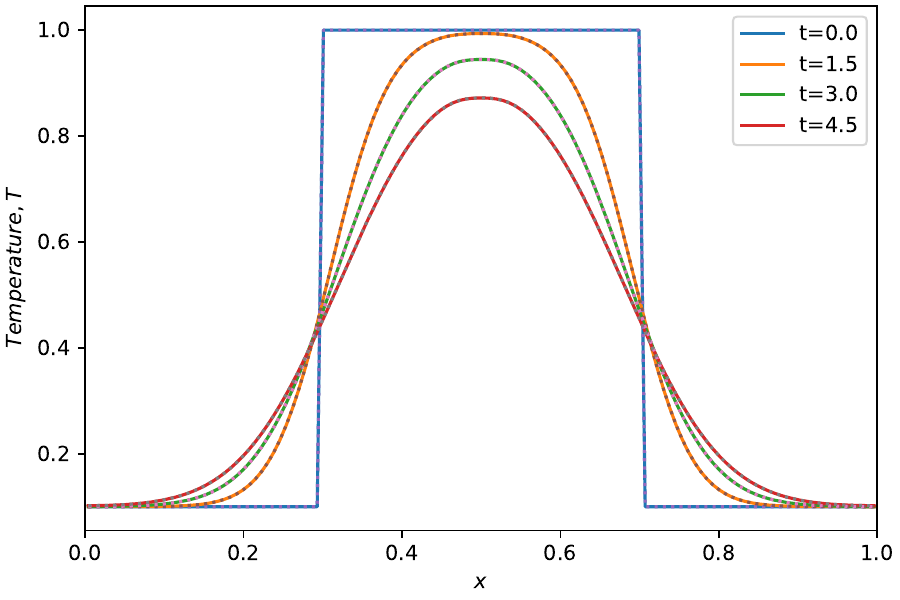}
    \caption{The widening of a top-hat temperature profile through dissipation by the heat flux within our toy model, both with (dotted lines) and without (continuous lines) the Chapman-Enskog expansion. There is excellent agreement between the two, with the small numerical errors being $\mathcal{O}(\tau_q)$.}
    \label{fig:toyq_tophat_compare}
\end{figure}

In \cref{fig:toyq_tophat_compare}, we use a `top-hat' initial temperature profile to compare the CE model~\cref{eq:toy_heat_CE_simple} to its originating equations~\cref{eq:toy_heat}. We see excellent agreement between the two. The performance improvement will be quantified below.

\subsection{Simple Viscosity Models}

\begin{figure}
    \centering
    \includegraphics[scale=0.58]{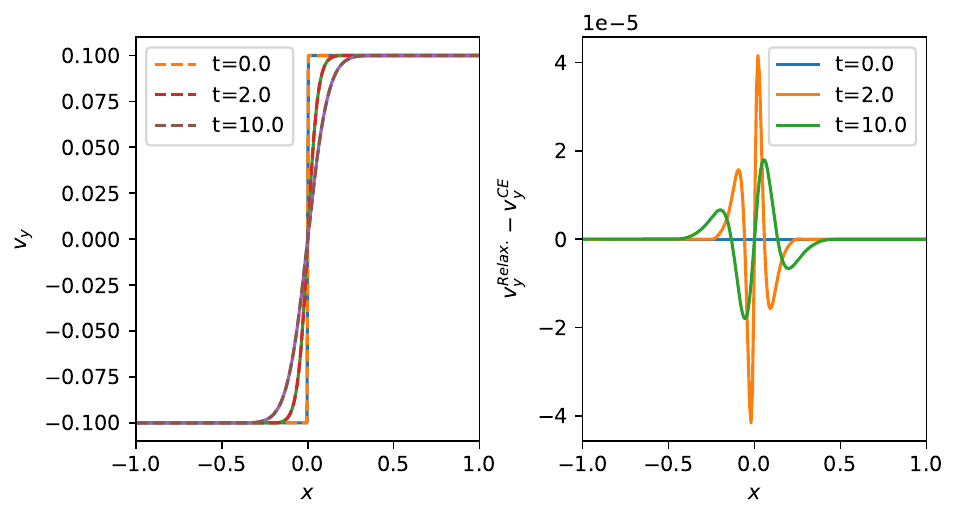}
    \caption{The evolution of the $y$-directed component of the velocity plotted across the $x$-domain at times $t=0.0, 2.0, 10.0$. The initial data for the velocity form a step function and the viscous parameter values are $\eta = \num{2e-4}$ and $\tau_\pi = \num{2e-4}$. In the left panel, two models' results are plotted: firstly, the MIS-derived simple shear model given by equations~\cref{eq:shear_flow_toy} (solid lines); secondly, the model obtained applying the CE-expansion to it, given by equation~\cref{eq:shear_flow_toy_CE} (dotted lines). In the right panel, the difference between the two results is plotted. The shear viscosity damps the initial step function, causing the velocity to develop approximately according to the analytic error-function. There is no visual difference seen between the two models. The numerical difference is an order of magnitude smaller than even the value of $\mathcal{O}(\tau_\pi)$ and is decreasing in time.}
    \label{fig:toy_pixy_shear_compare}
\end{figure}

Similar `toy' models that govern the evolution of bulk and shear viscosity may be derived from the MIS equations. For the bulk viscosity, in one dimension the relevant equations are 
\begin{subequations}
\label{eq:toy_bulk}
  \begin{align}
      \partial_t v_j + \partial_i \left( v^i v_j + \Pi \delta^i_j \right) &= 0, \label{eq:toy_bulk_v} \\
      \partial_t \Pi + \partial_i ( \Pi v^i ) &= -\frac{1}{\tau_{\Pi}} ( \zeta \partial_k v^k + \Pi),\label{eq:toy_bulk_pi}
  \end{align}
\end{subequations}
which reduces to the single 1D equation 
\begin{equation}
    \partial_t v +\partial_x (v^2) = \zeta \partial_{xx} v + \tau_{\Pi} \zeta \left(\partial_x v \partial_{xx} v - v \partial_{xxx} v \right)
    \label{eq:toy_bulk_CE}
\end{equation}
when the CE expansion is performed. For the shear viscous case, if we work in two spatial dimensions but consider a purely $y$-directed flow ($v_x = 0$) we have
\begin{subequations}
\label{eq:shear_flow_toy}
  \begin{align}
      \partial_t v_y + \partial_x  \pi_{xy} &= 0, \label{eq:shear_flow_toy_vy} \\
      \partial_t \pi_{xy} &= - \frac{1}{\tau_{\pi}} ( 2 \eta \sigma_{xy} + \pi_{xy}). \label{eq:shear_flow_toy_pi}
  \end{align}
\end{subequations}
Given that $\sigma_{xy}$ reduces to $\partial_x v_y$, we have
\begin{equation}
    \label{eq:shear_flow_toy_CE}
    \partial_t v_y = 2 \eta \partial_x^{(2)} v_y - 4 \eta^2 \tau_\pi \partial_x^{(4)}v_y
\end{equation}
as the CE form which governs the shear damping of the fluid velocity. This has an analytic solution in the limit $\tau_\pi \to 0$ given by $v_y(t,x) \sim v_y(0,x) \erf(\frac{x}{\eta t})$, which allows us to test convergence of the CE expression as the dissipation timescale vanishes. In \cref{fig:toy_pixy_shear_compare}, this test case is plotted, and again excellent  agreement is seen between the original relaxation model, the CE-expanded one.

Finally, \cref{tab:toy_Pi_runtimes} gives a comparison of run-times that indicates a significant speed-up is achievable using the CE form without compromising on accuracy. This speed-up of a factor between 2 and 3 comes purely from the reduced system size -- both models are evolved with the same, explicit time-integrators. This is possible for simple systems like this one, but is often not in the case of the MIS model, depending on the chosen parameter values. For the full MIS model where more expensive time integrators would be needed the speed-up will be larger, as investigated below.
\begin{table}
    \centering
    \begin{tabular}{c c|c c|c c}
    & & \multicolumn{2}{|c|}{\textbf{Model Runtimes} [s]} & & \\
         \textbf{Grid $N_x$} & \textbf{$t_{\text{final}}$} & Full & Chapman-Enskog & Scaling & Speed-up \\ \hline
         2048 & 1.0 & 5.40 & 1.60 & -- (--) & 3.4 \\ 
         4096 & 1.0 & 21.92 & 7.77 & 4.1 (4.9) & 2.8 \\ 
         8192 & 1.0 & 92.07 & 33.17 & 4.2 (4.3) & 2.9
    \end{tabular}
    \caption{Code run-times for our simple bulk viscosity model in one dimension, simulating the initial data seen in~\cref{fig:toy_pixy_shear_compare}. The expected scaling of the run-time, $t_{run} \propto N_x^2$, can be seen for both the full model and the Chapman-Enskog expansion (the latter in parentheses). Crucially, a nearly three-fold speed-up is achieved by using the Chapman-Enskog expansion, even with identical numerical methods.}
    \label{tab:toy_Pi_runtimes}
\end{table}

\section{General Balance-Law Derivation}
\label{sec:MISCE_Derivation}

We begin with the full MIS model in balance law form, \cref{eq:cons law}, recalling that the non-stiff and stiff conserved variables are labelled $\bm{q}$ and $\overline{\bm{q}}$ respectively. In order to maintain finite solutions in the ideal limit, we require that $\lim_{\tau \to 0} \overline{\bm{s}}(\bm{q}, \overline{\bm{q}})=\bm{0}$. This motivates an expansion of the stiff variables in powers of $\tau$, with each increasing order providing a further deviation from the ideal limit. 

In fact, because of the mathematical and physical links between the dissipation timescales ($\tau$) and strengths ($\xi$) discussed earlier, we choose to perform the expansion in powers of $\epsilon$ where both $\tau$ and $\xi$ are $\mathcal{O}(\epsilon)$. This reflects that in practice these parameters often take on similar (small) values in numerical simulations. This choice also means that $\mathcal{O}(\epsilon^0)$ corresponds to ideal behaviour with no dissipation. We could equally well perform the expansion in powers of $\tau$ and arrive at the same result, simply with a shifted series definition. Moving on, we now have
\begin{align}\label{CE expansion}
	\overline{\bm{q}} = \overline{\bm{q}}_0 + \overline{\bm{q}}_1 + \overline{\bm{q}}_2 + \mathcal{O}(\epsilon^3)
\end{align}
where $\overline{\bm{q}}_0$ is $\mathcal{O}(\epsilon^0)$, $\overline{\bm{q}}_1$ is $\mathcal{O}(\epsilon)$, $\overline{\bm{q}}_2$ is $\mathcal{O}(\epsilon^2)$ and so on. To identify the terms in this expansion we take the form of the stiff source:
\begin{align}\label{eq:stiff source}
	\overline{\bm{s}}(\bm{q}, \overline{\bm{q}}) = \frac{1}{\epsilon}(\overline{\bm{q}} - \overline{\bm{q}}_{NS})
\end{align}
and simply rewrite it as 
\begin{align}\label{eq:stiff source linear}
	\epsilon \overline{\bm{s}}(\bm{q}, \overline{\bm{q}}) = \overline{\bm{q}} - \overline{\bm{q}}_{NS}.
\end{align}	
Noting that $\bm{q}_{NS}$ is $\mathcal{O}(\epsilon)$, we have that at zero'th order (the ideal case) $\overline{\bm{q}} = \overline{\bm{q}}_0 = \bm{0}$. At first order we have $\overline{\bm{q}} = \overline{\bm{q}}_1 = \overline{\bm{q}}_{NS}$ and at second order we have $\overline{\bm{q}} = \overline{\bm{q}}_2$ where $\overline{\bm{q}}_2$ is yet to be determined.

At zeroth-order, the non-stiff subsystem of equations is given by
\begin{align}\label{eq:cons law ZO}
        \partial_t {\bm{q}}_0(\bm{w}) + \partial_i \bm{f}_0^{i}(\bm{w}) = \bm{0}
\end{align}
where
\begin{equation}\label{eq:cons law ZO terms}
      {\bm{q}}_0(\bm{w}) = \begin{pmatrix}
      D \\ S_j \\ \tau \end{pmatrix} = \begin{pmatrix}
      n W \\ (\rho + p) W^2 v_j \\ (\rho + p) W^2 - p - nW \end{pmatrix}
\end{equation}
and      
\begin{equation}
      \bm{f}_0^{i}(\bm{w}) = \begin{pmatrix}
      nW v^i \\ (\rho + p) W^2 v^i v_j + p \delta^i_j \\ (\rho + p) W^2 v^i - nW v^i \end{pmatrix}. \\   
\end{equation}
These are simply the relativistic Euler equations. At first-order, it can be written as 
\begin{equation}
    \begin{split}
        \label{cons law FO}
        \partial_t \left[ \bm{q}_0(\bm{w}) + \bm{H}_{(1)}(\bm{w}, \partial_t\bm{w}, \partial_i\bm{w}) \right] + \\  \partial_i \left[ \bm{f}_0^{i}(\bm{w}) + \bm{F}_{(1)}^{i}(\bm{w}, \partial_t\bm{w}, \partial_i\bm{w}) \right] = 0
    \end{split}
\end{equation}
where
\begin{equation}
        \bm{H}_{(1)}(\bm{w}) =
        \begin{pmatrix} 0 \\ W(q_{0,NS} v_j + q_{j,NS}) + \pi_{0j,NS} \\ 2q_{0,NS} W - \Pi_{NS} + \pi_{00,NS}) \end{pmatrix} 
\end{equation}
and 
\begin{equation}
        \bm{F}_{(1)}^{i}(\bm{w}) = 
        \begin{pmatrix} 0 \\ \Pi_{NS} (v^i v_j + \delta^i_j)W^2  + W(q^i_{NS} v_j + q_{j,NS} v^i) + \pi^i_{j,NS} \\ W(q_{0,NS} v^i + q^i_{NS}) + \pi_{0,NS}^i \end{pmatrix}. 
\end{equation}
Here, we have separated the dissipative parts of the state and flux vectors and can view $\bm{H}_{(1)}$ and $\bm{F}_{(1)}$ as $\mathcal{O}(\epsilon)$ perturbations on-top of the ideal $\mathcal{O}(\epsilon^0)$ state and flux vectors, $\bm{q}_0$ and $\bm{f}_0$. In general, we can rewrite the expanded system as
\begin{align}\label{eq:cons law CE expansion}
    \partial_t \bm{q}_0 + \partial_i \bm{f}_0^{i} = \sum_{p=0} \tilde{\bm{R}}_{(p)} \equiv \sum_{p=0} \left( -\partial_t \bm{H}_{(p)} - \partial_i \bm{F}^{i}_{(p)} \right)
\end{align}
where each additional term in the series on the RHS of \cref{eq:cons law CE expansion} represents a source correction of order $\epsilon^p$. Hence, $\tilde{\bm{R}}_{(0)} = \bm{0}$, $\tilde{\bm{R}}_{(1)} = -\partial_t \bm{H}_{(1)} - \partial_i \bm{F}_{(1)}^{i}$ and $\tilde{\bm{R}}_{(2)} = -\partial_t \bm{H}_{(2)} - \partial_i \bm{F}_{(2)}^{i}$ and so on. 

Using symbolic Python, we have fully derived the first-order ($\mathcal{O}(\epsilon)$) source terms in $\tilde{\bm{R}}_{(1)}$ such that they contain only spatial gradients. At second order, we have derived the flux contribution to $\tilde{\bm{R}}_{(2)}$ that is $- \partial_i \bm{F}_{(2)}^{i}$. The presence of high order time derivatives in $-\partial_t \bm{H}_{(2)}$, which in turn introduce even higher-order spatial derivatives, leads to algebraic terms that rapidly scale in number and complexity, making it impractical to derive and implement, even using computer algebra packages. 

Note that one cannot always directly align powers of $\epsilon$ ($\{\zeta, \kappa, \eta\}$ or $\{ \tau_\Pi, \tau_q, \tau_\pi$\}) with the order of spatial derivatives appearing in these source terms. To see this, consider the simple (CE) bulk viscosity and heat flux models from earlier given by~\cref{eq:toy_heat_CE_simple} and~\cref{eq:toy_bulk_CE}. In the former, the next-to-leading order correction is $\mathcal{O}(\kappa^2 \tau_q$) and contains a fourth-order derivative, whilst in the latter it is $\mathcal{O}(\zeta \tau_\Pi$) and contains a mixture of first-, second- and third-order derivatives. However, the leading order correction in each case, $\tilde{\bm{R}}_{(1)}$, contains mostly second-order gradients in the primitive variables ($\partial_i \partial_j {\bm w}$) with some products of two first-order derivatives ($\partial_i {\bm w} \partial_j {\bm w}$). To see this, consider that the dissipative variables we move from the state and flux vectors to the new sources contain first order gradients. If moved from the flux vector, becoming $-\partial_i \bm{F}_{(1)}^{i}$, they pick up another spatial derivative from the flux-gradient. If moved from the state vector, becoming $-\partial_t \bm{H}_{(1)}$, they pick up a first-order temporal derivative, which we will show can be swapped for a first-order spatial derivative. Hence, they are always diffusive, second-order gradients as one would expect for dissipation. This can also be seen in ~\cref{eq:toy_heat_CE_simple} and~\cref{eq:toy_bulk_CE} at leading-order. 

We choose to perform the series expansion and truncation such that terms $\mathcal{O}(\epsilon)$ contain no timescales and are first-order in the dissipation strengths $\{\zeta, \kappa, \eta\}$. Terms considered to be $\mathcal{O}(\epsilon^2)$ are first-order in the timescales $\{ \tau_\Pi, \tau_q, \tau_\pi$\} and the strengths. We often choose to work with the first order ($\mathcal{O}(\epsilon)$) source terms only as we find that including higher orders generally only makes small quantitative differences. However, using the $\mathcal{O}(\epsilon^2)$ source, the effect of varying timescales for both the MIS and MISCE models will be shown. Finally, despite the inherent instability of first-order theories of relativistic dissipation in fluids~\citep{hiscock_stability_1983}, we do not find any instabilities arising with our first-order MISCE model, at least for the test problems and parameter space explored so far.

In order to make it practical to implement the system numerically, we need to replace the time derivatives present in $\tilde{\bm{R}}_{(1)}$ and $\tilde{\bm{R}}_{(2)}$ with spatial ones. We have two potentially problematic sources of time-derivatives. Firstly, the Navier-Stokes forms of the dissipative variables themselves contain time derivatives. Secondly, the entire dissipative state vector $\bm{H}(\bm{w})$ is time-differentiated in the equations of motion. Because both $\bm{H}$ and $\bm{F}$ can be expressed entirely as functions of primitive, non-stiff variables, we need expressions for the time derivatives of the primitive variables. Making use of the chain rule and \cref{eq:cons law ZO}, which contains the time derivative of the ideal state vector and hence the primitive variables that constitute it, we have
\begin{equation}\label{eq:dwdt expansion}
    \frac{\partial \bm{w}}{\partial t} = \frac{\partial \bm{w}}{\partial \bm{q}_0} \frac{\partial \bm{q}_0}{\partial t} + \frac{\partial \bm{w}}{\partial \bm{q}_1} \frac{\partial \bm{q}_1}{\partial t} + ... = -\left(\frac{\partial \bm{q}_0}{\partial \bm{w}}\right)^{-1} \partial_i\bm{f}^{i}_0 + \mathcal{O}(\epsilon),
\end{equation}
where we again note that $\bm{w}$ is the vector of primitives. This means the term $\frac{\partial \bm{w}}{\partial \bm{q}_0}$ has a matrix form that is far more easily obtained through an inversion of the matrix $\frac{\partial \bm{q}_0}{\partial \bm{w}}$. We can use this result to substitute wherever a time-derivative appears in our source such that we then have
\begin{equation}\label{eq:R1}
    \tilde{\bm{R}}_{(1)} = -\partial_t \bm{H}_\text{NS}(\bm{w},\partial_i\bm{w}) - \partial_i\bm{F}_\text{NS}^{i}(\bm{w},\partial_i\bm{w})
\end{equation}
and our source contains solely first and second-order \emph{spatial} derivatives. For a derivation of higher-order approximations to time derivatives of primitive variables, see~\cref{app:dwdt_NLO}.

We will dub this new formulation DEIFY (Dissipative Extension to Ideal Fluid dYnamics) so that $\tilde{\bm{R}}_{(1)}$ is the first-order DEIFY source term. Also observe how the source term for DEIFY is proportional to $\epsilon$ whereas the M{\"u}ller-Israel-Stewart formulation source terms scale as $1/\tau \propto \epsilon^{-1}$. This means that the two forms become stiff in opposing limits---near the ideal regime (small $\epsilon$) DEIFY will be stable as a result of a small source term, and will only become stiff, and potentially unstable, as $\epsilon$ grows large. The big benefit of this behaviour is that near the ideal regime we can confidently evolve DEIFY with explicit time integrators, knowing that source contributions will remain small.

In contrast, in the event of very slow-acting (large $\tau$) and large-in-magnitude (large $\{\kappa, \zeta, \eta\}$) viscosities and heat fluxes, it will not be sensible or accurate to evolve DEIFY, even using implicit schemes. Instead, we may revert to an implementation of the MIS formulation in this regime, which is likely to be stable with explicit integrators and therefore less costly. Future work will extend the approach of~\cite{wright_non-ideal_2020}, where an adaptive model of resistive and ideal MHD was implemented. Ours will be able to switch between different dissipative formulations of hydrodynamics during evolution, ensuring stability, efficiency and accuracy.

In summary, in both the ideal and highly-non-ideal limits, we should be able to use explicit integration schemes, which have been shown to provide a speed-up of up to an order of magnitude over implicit schemes in comparable models of resistive/ideal MHD~\citep{wright_resistive_2020}. In \cref{subsec:code_performance} we will prove the validity of this claim, and further explore the intermediate region of non-ideal behaviour between these two extremes.

\subsection{First-Order Source}

In order to compute the DEIFY source term(s), we will need to calculate matrices and, crucially, their inverses. For instance, we will need to know the inverse matrix that appears on the RHS of \cref{eq:dwdt expansion}, that represents the Jacobian of the primitive vector with respect to the non-stiff conserved vector.

Here, we have a choice of how to compute the matrices of interest---that is we can invert them numerically, or try to get the form of the inverted matrix symbolically. Inverting matrices numerically, especially when densely populated, can require a large amount of computation, reducing accuracy as well as slowing down simulations. If the algebraic form of the matrices were at hand, this would lead to a far more efficient simulation, and as we are trying to build a source term to extend ideal hydrodynamics with the intention of being faster to evolve than other forms of dissipative hydrodynamics, it is sensible to adopt the performance gains of a purely symbolic source term. 

On this note, let us turn to computing (algebraically) the matrices $\left(\frac{\partial \bm{q}_0}{\partial \bm{w}}\right)$ and, hence, $\left(\frac{\partial \bm{q}_0}{\partial \bm{w}}\right)^{-1}$. Here, we will make a simplification so that the terms appearing in these matrices are human-readable: we take the low-velocity limit, neglecting terms $\mathcal{O}(v^2)$ and hence setting the Lorentz factor, $W = 1$. This assumption is not made for the numerical implementation. We also have a choice to make over which two thermodynamic variables are present in our primitive variable vector $\bm{w}$. The equation of state, which relates $p$, $\rho$ and $n$, gives us this choice, and we opt to work with $\bm{w} = \{p, \rho, \bm{v}\}$. Thus, our ideal conserved vector is now given by
\begin{align}\label{eq:cons law ZO terms low-vel}
      {\bm{q}}_0(\bm{w}) &= \begin{pmatrix}
      D \\ S_j \\ E = \tau + D \end{pmatrix} = \begin{pmatrix}
      \rho + p/(1-\Gamma) \\ (\rho + p) v_j \\ \rho \end{pmatrix}
\end{align}
where we have used our equation of state to replace $n$ in the expression for $D$ and chosen to work with the conserved variable $E$ for now instead of $\tau$ as it takes a simpler form. 

This gives us the matrix
\begin{align}\label{eq:dq0dw}
    \left(\frac{\partial \bm{q}_0}{\partial \bm{w}}\right) = \begin{pmatrix} \partial_p n & \partial_\rho n & 0 & 0 & 0 \\
    v_1 & v_1 & p +\rho & 0 & 0 \\
    v_2 & v_2 & 0 & p +\rho & 0 \\
    v_3 & v_3 & 0 & 0 & p +\rho \\
    0 & 1 & 0 & 0 & 0 \end{pmatrix}
\end{align}
and, hence, $\left(\frac{\partial \bm{q}_0}{\partial \bm{w}}\right)^{-1}$ is
\begin{align}\label{eq:dq0dw-1} 
(p+\rho)^{-1} \begin{pmatrix} (p+\rho)/\partial_p n & 0 & 0 & 0 & -(p+\rho)\partial_\rho n / \partial_p n \\
    0 & 0 & 0 & 0 & (p+\rho) \\
    -v_1/\partial_p n  & 1 & 0 & 0 & -v_1(\partial_p n + \partial_\rho n) / \partial_p n \\
    -v_2/\partial_p n & 0 & 1 & 0 & -v_2(\partial_p n + \partial_\rho n) / \partial_p n \\
    -v_3/\partial_p n & 0 & 0 & 1 & -v_3(\partial_p n + \partial_\rho n) / \partial_p n \end{pmatrix}.
\end{align}
Next, using \cref{eq:dwdt expansion}, this gives us
\begin{align}\label{eq:dwdt in Cons}
    \partial_t \begin{pmatrix} p \\ \rho \\ v_j \end{pmatrix}
    = (\rho + p)^{-1} \begin{pmatrix} (p+\rho)( (1 / \partial_p n)  - (\partial_\rho n / \partial_p n )\partial_t E) \\ (p+\rho) \partial_t E \\ -v_j((1+\partial_\rho n / \partial_p n)  \partial_t E + (1/\partial_p n)\partial_t D) + \partial_t S_j \end{pmatrix}
\end{align}
where we can exchange the time-derivatives of the conserved variables for spatial derivatives of the fluxes using \cref{eq:cons law ZO}. Doing this, and using our equation of state $p = (\Gamma - 1)(\rho - n)$ to replace the partial derivatives of primitive variables, we arrive at 
\begin{align}\label{eq:dwdt in Cons_gamma_law}
    \partial_t \begin{pmatrix} p \\ \rho \\ v_j \end{pmatrix}
    = \begin{pmatrix} (1 - \Gamma)(1 + \partial_i S^i) \\ -\partial_i S^i \\ (\rho + p)^{-1} \left[v_j((2 - \Gamma)  \partial_i S^i + (1-\Gamma)\partial_i (Dv^i)) - \partial_i S^i_j \right] \end{pmatrix}
\end{align}
which represents expressions for the partial time derivatives of the primitive variables in terms of purely spatial-derivatives. 

Let us now demonstrate what the MISCE sources look like. These are too complex to write in full, so we consider the case of bulk viscosity only, restrict to one spatial dimension and again work in the low-velocity approximation such that $W = 1$ (but not neglecting terms $\mathcal{O}(v^2)$). Then, $\Pi_{NS} = -\zeta \partial_x v^x$. The leading order source term, $\tilde{\bm{R}}_{(1)} = -\partial_t \bm{H}_{(1)} - \partial_x \bm{F}_{(1)}^{x}$ in full is
\begin{subequations}
  \begin{align}
    \tilde{\bm{R}}_{(1)} &= -\partial_t \begin{pmatrix} 0 \\ 0 \\ -\Pi_{NS} \end{pmatrix} - \partial_x \begin{pmatrix} 0 \\ \Pi_{NS} v_x^2 \\ 0 \end{pmatrix} \\
    &= \zeta \left[  \begin{pmatrix} 0 \\ 0 \\ -\partial_t \partial_x v^x \end{pmatrix} + \partial_x \begin{pmatrix} 0 \\ v_x^2 \partial_{xx} v^x + 2 v_x (\partial_x v^x)^2 \\ 0 \end{pmatrix} \right]
  \end{align}
\end{subequations}
where from~\cref{eq:dwdt in Cons_gamma_law} we have
\begin{subequations}
  \begin{align}
    \partial_x \partial_x v^t &=  (\rho + p)^{-1} [ ((2 - \Gamma) (v_x \partial_{xx} S^x + (\partial_x S^x )(\partial_x v_x )) \notag \\ &+ (1-\Gamma)\partial_{xx} (D v_x)) - \partial_{xx} (S^x v_x + p) ] \\
    &= [ ((2 - \Gamma) (2v_x \partial_{xx} v_x + 3 v_x (\partial_x v_x) \partial_x (\rho + p) + (\partial_x v_x)^2 ) \notag \\ &+ (\rho + p)^{-1} (1-\Gamma) (v_x \partial_{xx} n + n \partial_{xx} v_x ) \notag \\
    &- \partial_{xx} v_x -  v_x \partial_x (\rho + p) - (\rho + p)^{-1} \partial_{xx} p) ]
  \end{align}
\end{subequations}
and, finally, we have an expression for $\tilde{\bm{R}}_{(1)}$ that is expressed purely in (second-order) spatial gradients of the primitive variables. The full expressions (without simplification) are not human-readable, but the code to derive them can be found at \href{CompAlg}{https://www.github.com/MarcusHatton/ComputerAlgebra}, whilst their implementation can be seen at \href{MyMETHODFork}{https://www.github.com/MarcusHatton/METHOD}.

\subsection{Second-Order Source}

A similar but more complex calculation can be made to derive the next order (second) of dissipative correction to ideal hydrodynamics. Beginning again with the conservation law
\begin{align}\label{eq:cons law SO}
        \partial_t \left[ \bm{q}_0(\bm{w}) + \bm{H}_{(1)} + \bm{H}_{(2)} \right] + \partial_i \left[ \bm{f}_0^{i}(\bm{w}) + \bm{F}^{i}_{(1)} + \bm{F}_{(2)}^{i} \right] = 0
\end{align}
it follows that
\begin{equation}
    \tilde{\bm{R}}_{(2)} = -\partial_t \bm{H}_{(2)} - \partial_i\bm{F}^{(i)}_{2}.
\end{equation}
At first order, the form of $\bm{H}$, $\bm{F}$ and hence $\tilde{\bm{R}}_{(1)}$ followed simply from the definition of the Navier-Stokes terms which are of $\mathcal{O}(\epsilon)$. At second order we use the stiff subsystem
\begin{subequations}
\begin{align}
	\partial_t \left( nW \overline{\bm{w}} \right) + \partial_i \left( nWv^i \overline{\bm{w}} \right) &= \frac{n}{\tau}(\overline{\bm{w}} - \overline{\bm{w}}_{NS}) \\
	\partial_t \overline{\bm{q}} + \partial_i \overline{\bm{f}}^i(\bm{q}, \overline{\bm{q}}) &= \frac{1}{\epsilon}(\overline{\bm{q}} - \overline{\bm{q}}_{NS})
\end{align}
\end{subequations}
and make the substitution $\overline{\bm{q}} = \overline{\bm{q}}_0 + \overline{\bm{q}}_1 + \overline{\bm{q}}_2 \equiv \bm{0} + \overline{\bm{q}}_{NS} + \overline{\bm{q}}_2$ to obtain, at order $\mathcal{O}(\epsilon^2)$,
\begin{align}
	\epsilon \left[ \partial_t \overline{\bm{q}}_{NS} + \partial_i \overline{\bm{f}}^i(\bm{q}, \overline{\bm{q}}_{NS}) \right] = \overline{\bm{q}}_2.
\end{align}

Because the NS forms of the stiff variables can be defined entirely in terms of the non-stiff primitive variables $\left( \overline{\bm{q}}_{NS} \equiv \overline{\bm{q}}_{NS}(\bm{q}) \right)$, so too can $\overline{\bm{q}}_2$. The vectors $\bm{H}_{(2)}$, $\bm{F}_{(2)}$ are given by 
\begin{equation}
        \bm{H}_{(2)}(\bm{w}) = 
        \begin{pmatrix} 0 \\ W(q_{0,(2)} v_j + q_{j,(2)}) + \pi_{0j,(2)} \\ 2q_{0,(2)} W - \Pi_{(2)} + \pi_{00,(2)}) \end{pmatrix} \notag
\end{equation}
and 
\begin{equation}
        \bm{F}^{i}_{(2)}(\bm{w}) = 
        \begin{pmatrix} 0 \\ \Pi_{(2)} (v^i v_j + \delta^i_j)W^2  + W(q^i_{(2)} v_j + q_{j,(2)} v^i) + \pi^i_{j,(2)} \\ W(q_{0,(2)} v^i + q^i_{(2)}) + \pi_{0,(2)}^i \end{pmatrix}. \notag
\end{equation}

Putting these results together and making substitutions wherever we find time-derivatives of the primitive variables (as before) allows us to arrive at a purely spatial form for $\tilde{\bm{R}}_{(2)}$.

\section{Results}
\label{sec:results}

Next, we will perform numerical tests of our implementation of the MISCE model, with comparison primarily to the usual MIS model from which it has been derived. A code named Multifluid Electromagneto-HydroDynamics (METHOD) was used to perform these simulations, which may be found at \href{My METHOD fork}{https://www.github.com/MarcusHatton/METHOD}, having been forked and extended from its creator's repository at \href{Alex's METHOD Repo}{https://www.github.com/AlexJamesWright/METHOD}. Instructions on how to run the simulations and reproduce the results of this chapter are to be found on the `MISCE Paper' branch.

These will be standard tests such that we may compare results against the literature and check for agreement. These tests are also chosen to reflect the physics we are interested in capturing for actual neutron star mergers. 

\subsection{Shocktubes}

Shocktubes are simple, one-dimensional tests useful for closely analysing the behaviour of the fluid model and its numerical implementation. They are designed to produce a set of forward- and backward-travelling waves, and in particular, discontinuities in the fluid's properties. These waves (contact, rarefaction and shock) are the fundamental propagation modes of the fluid will certainly be produced at the point of merger, and thereafter whenever a sharp jump in density, pressure or temperature  occurs such as between different phases of matter within the neutron star. They also involve advection of the fluid, which will be important for the inspiral phase of the merger as tidal forces will drag fluid around the star as they orbit their mutual centre of mass.

The initial data for these tests are similar to those of~\cite{takamoto_fast_2011}. We also share the same equation of state, allowing for a favourable quantitative comparison to be made. A domain of one spatial dimension is initially split into bordering left and right states $[L,R]$ where $x \in x_L = [-1.0,0.0)$ and $x \in x_R = [0.0,1.0]$. The primitive variables in the two states are
\begin{equation}
    L: ~
    \begin{pmatrix}
        v_x \\
        p \\
        n \\
        \rho 
    \end{pmatrix} = 
    \begin{pmatrix}
        +0.2 \\
        10 \\ 
        10 \\
        25
    \end{pmatrix}
\end{equation}
for the left state and
\begin{equation}
    R: ~ 
    \begin{pmatrix}
        v_x \\
        p \\
        n \\
        \rho 
    \end{pmatrix} = 
    \begin{pmatrix}
        -0.2 \\
        1 \\ 
        1 \\
        2.5
    \end{pmatrix}
\end{equation}
for the right state. The pressure, $p$, and number density, $n$, are set in the initial data and the equation of state, $p = (\Gamma -1)(\rho -n)$, determines the energy density, $\rho$. We use a value of $\Gamma = 5/3$ throughout these simulations.

\begin{figure}
    \centering
    \includegraphics[scale=0.475]{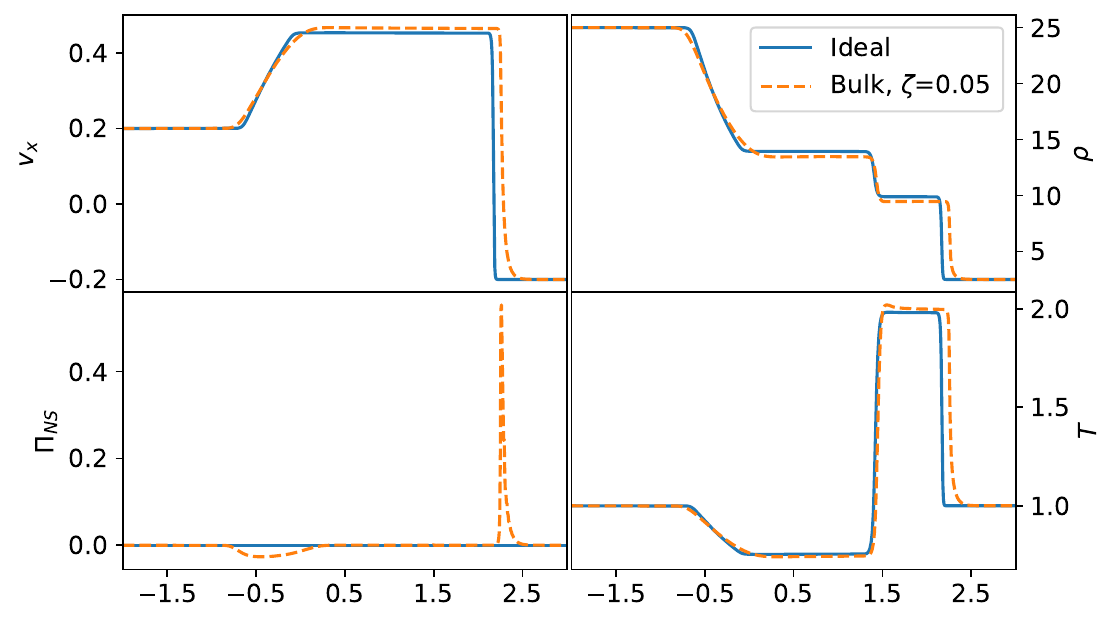}
    \caption{A shocktube simulation comparing results of an ideal fluid and one with a bulk viscosity parameter of $\zeta = \num{5e-2}$, using the MISCE formulation. The velocity, energy density, Navier-Stokes bulk viscosity and temperature are plotted. The shift in shock propagation speed, damping of the energy density, and additional heating due to the inclusion of bulk viscosity are all visible effects.}
    \label{fig:ST_ISCE_IdvsBulk_nvxrhoPiNS}
\end{figure}

In~\cref{fig:ST_ISCE_IdvsBulk_nvxrhoPiNS}, we see the expected production of the three travelling waves: the left-moving rarefaction; right-moving contact-wave; and (faster) right-moving shockwave. These are most easily seen in the energy density plot in the top-right. The bulk viscosity has a smoothing effect on these waves, particularly on the shockwave, where it also significantly increases the propagation speed of the shock-front -- this can also be seen directly in the velocity plot (top-left). The bulk viscosity itself (bottom-right) spikes at the shock where the velocity gradients are highest. Its positivity there indicates a resistance to the rapid compression of the fluid by the shockwave, with the reverse being true for the rarefaction. Finally, the wide temperature peak between the contact-wave and shockwave is magnified slightly by the inclusion of bulk viscosity, as is to be expected from such a dissipative effect.

\begin{figure*}
    \centering
    \includegraphics[width=1.0\linewidth]{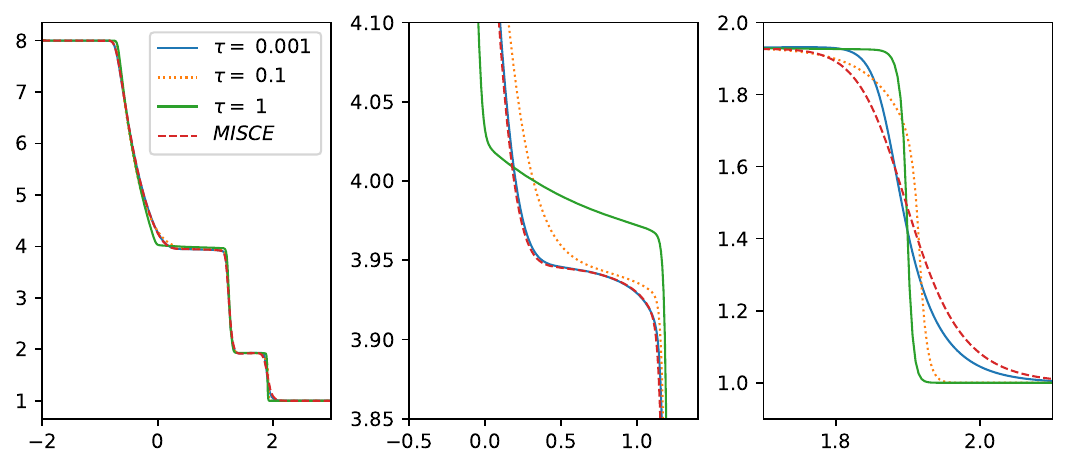}
    \caption{The evolution of the number density for a `stillshock' test--a shocktube with zero initial velocity. There is bulk viscosity and heat flux present with coefficients $\zeta = \num{5e-2}$ and $\kappa = \num{5e-3}$. The three panels show the entire domain (left), the rarefaction wave (centre) and the shockwave (right). The two models (MIS, MISCE) are compared in all three panels, with the dissipative timescale $\tau$ varying for the MIS model but held constant at zero for the MISCE model. One can see the approach of the MIS solution towards the MISCE solution as $\tau \to 0$. For the rarefaction wave they have converged in the fastest case, but for the shock there are still differences. In particular, one can see the increase in speed of the shock as the ratio $\zeta / \tau_\pi$ increases for the MIS model. It is catching up to the MISCE solution, which can be thought of as its limiting case.}
    \label{fig:SS_MISvsMISCE_Bulk_Zooms}
\end{figure*}

In~\cref{fig:SS_MISvsMISCE_Bulk_Zooms}, we take a closer look at a shocktube profile for the fluid's number density with bulk viscosity and heat flux present. Three results from the MIS model are plotted for differing dissipative timescales $\tau$ (the same for both types of dissipation), and one for the MISCE model (at leading order, so the timescale does not enter into the EoM). In particular, we see convergence of the MIS result to the MISCE result as $\tau$ is decreased. This is expected given that for the MIS model, in the $\tau \to 0$ limit, any off-shell deviations from relativistic Navier-Stokes behaviour are instantaneously quenched. This means that the MIS model's behaviour should match that of the leading-order MISCE model in this limit, where terms $\mathcal{O}(\tau)$ and higher are neglected.


\subsection{Kelvin-Helmholtz Instabilities}

The Kelvin-Helmholtz instability (KHI) is a shearing instability that results when two (or more) fluid regions flow in opposite directions past each other, each usually differing in density. A wide range of fluid behaviours can be observed depending on the precise initial data, but here we will be varying the shear viscosity only, to focus on its effect. In the most interesting cases, there is an initial linear growth phase of the instability at the interface, followed by a non-linear phase where the creation of vortices and a complex network of shocks typically precedes the onset of smaller-scale turbulence. 

Neutron star matter in mergers is likely to be Kelvin-Helmholtz unstable as the two objects collide and shearing flows develop. The KHI is known to play an important role in post-merger dynamics where it moderates the cascade of energy between macroscopic and microscopic scales through the action of shear viscosity in the fluid. This is important in the spin-down of the remnant where the rotational energy of the fluid is converted to small-scale turbulence and then to either magnetic energy through the dynamo effect or dissipated through viscous heating. We will also analyse the integrated power spectrum of kinetic energy resulting from turbulence induced by the KHI. This has famously been shown by Kolmogorov to have a universal scaling relation with wavenumber for at least part of its spectrum, a result which was generalized by~\cite{qian_generalization_1994}.

To investigate this process, we use the initial conditions from~\cite{beckwith_second_2011}, as well as the spectral analysis laid out by them. The data are defined within a 2D domain where $x \in [-1.0,1.0]$ and $y \in [-0.5,0.5]$. The domain is then divided into two fluid regions, with the inner region contained roughly within $x \in [-0.5,0.5]$ and the outer elsewhere. The two fluid regions have differing densities and flow past each other with velocities directed in the positive and negative $y$-directions. There is a narrow transition layer between the two where a small, spatially-varying perturbation to the $x$-directed velocity is also introduced to induce mixing. The primitive variables are
\begin{subequations}
    \begin{align}
    \begin{pmatrix}
        v_y \\
        \rho \\
        v_x  
    \end{pmatrix} &=      \begin{pmatrix}
        v_{sh} \tanh\left(\frac{x-0.5}{a}\right) \\
        \rho_0 + \rho_1 \tanh\left(\frac{x-0.5}{a}\right) \\
        A_0 v_{sh} \sin(2\pi y)\exp\left(\frac{-(x-0.5)^2}{l^2} \right)   
    \end{pmatrix} ~ ; ~ x > 0.0
        \\
        \intertext{and}
    \begin{pmatrix}
        v_y \\
        \rho \\
        v_x  
    \end{pmatrix} &=      \begin{pmatrix}
        -v_{sh} \tanh\left(\frac{x+0.5}{a}\right) \\
        \rho_0 - \rho_1 \tanh\left(\frac{x+0.5}{a}\right) \\
        -A_0 v_{sh} \sin(2\pi y)\exp\left(\frac{-(x+0.5)^2}{l^2} \right)   
    \end{pmatrix} ~ ; ~ x \leq 0.0
    \end{align}
\end{subequations}

where the shear velocity is $v_{sh} = 0.5$, the boundary layer thickness is $a = 0.01$, the densities are given by $(\rho_0, \rho_1) = (0.55, 0.45)$, and the perturbation has an amplitude $A_0 = 0.1$ over a characteristic length $l = 0.1$. The initial pressure is uniform, $p = 1.0$, and the adiabatic index is set to $\Gamma = 4/3$. We use periodic boundaries in the both the $x$ and $y$ directions. 
\begin{figure}
    \centering
    \includegraphics[scale=1.05]{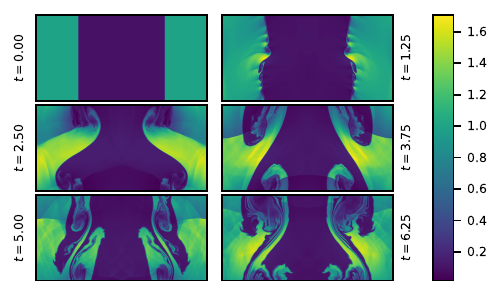}
    \caption{The development of a Kelvin-Helmholtz unstable fluid with negligible viscosity until $t=6.25$. The number density is shown in colour, as is the case for all KHI plots here. The initial perturbation grows rapidly until the interface breaks and large-scaling mixing occurs, followed by the onset of turbulent behaviour which produces shocks and smaller-scale vortices.}
    \label{fig:KH_MISCE_Id_800x1600}
\end{figure}
\begin{figure}
    \centering
    \includegraphics[scale=1.05]{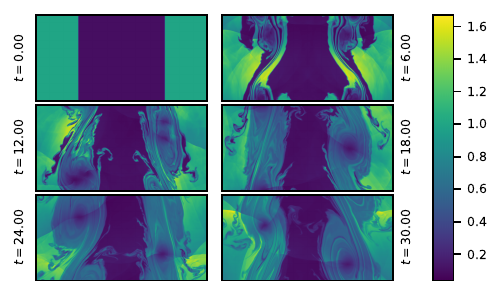}
    \caption{The development of the Kelvin-Helmholtz instability until $t=30.0$, for an ideal fluid with negligible viscosity. The longer simulation time allows the asymmetric initial perturbation at the interface to give rise to large-scale asymmetric vortex formation.}
    \label{fig:KH_MISCE_Id_t30_800x1600}
\end{figure}
\begin{figure}
    \centering
    \includegraphics[scale=1.05]{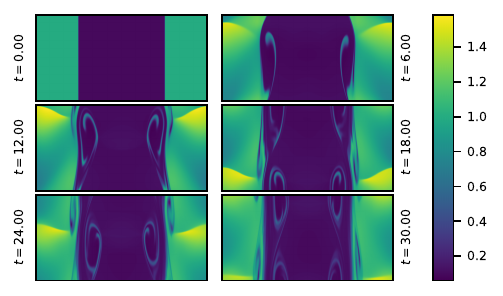}
    \caption{The long-term evolution of the Kelvin Helmholtz instability using the MISCE model at leading order with a shear viscosity parameter of $\eta = \num{1e-3}$. The shear viscosity has an intermediate value here: it suppresses large-scale mixing of the two fluids but vortices still form in a narrow shearing layer that is stable even at late times. The asymmetry is again visible here, but obscured for similar reasons.}
    \label{fig:KH_MISCE_Sh}
\end{figure}

\Cref{fig:KH_MISCE_Id_800x1600,fig:KH_MISCE_Id_t30_800x1600,fig:KH_MISCE_Sh} show the development of the KHI for the fluid's number density. \Cref{fig:KH_MISCE_Id_800x1600,fig:KH_MISCE_Id_t30_800x1600} show its development for an ideal (inviscid) fluid. For the former, the early-time behaviour is the focus, with the initial growth of the interface instability visible, followed by large-scale mixing and finally the formation of small-scale structure as energy cascades from longer to shorter scales. In the latter, the asymmetry of the initial perturbation has had time to grow into a macroscopic asymmetry. One can also see vortices forming and the onset of turbulence in the wide mixing layer.

In \cref{fig:KH_MISCE_Sh}, the long-term behaviour for a viscous fluid is shown. Viscosity suppresses the perturbation's growth and stabilises the mixing at the interface. Vortices do form, still, but they are confined to a smaller corridor between the two bulk fluid regions, and in general the behaviour is less chaotic. We observe similar qualitative behaviour to~\cite{takamoto_fast_2011}, who performed comparable simulations. 
\begin{figure}
    \centering
    \includegraphics[scale=0.55]{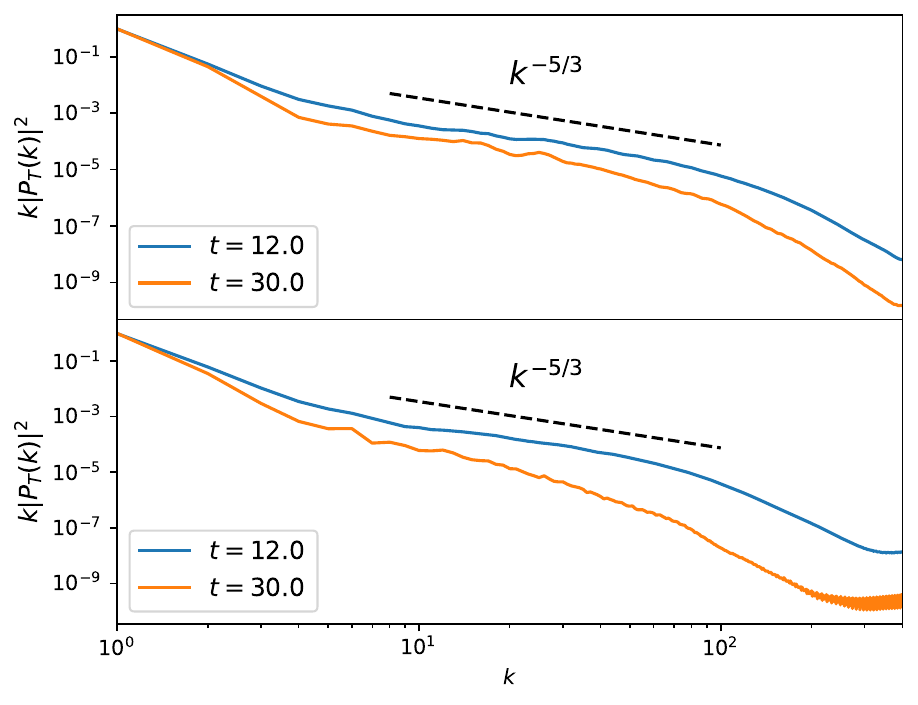}
    \caption{The power spectra for the kinetic energy density in the Kelvin-Helmholtz instability at medium and long times, for an inviscid fluid (top) and one with weak shear viscosity, $\eta = \num{1e-4}$ (bottom). This uses the MISCE formulation with a grid of size $N_x = N_y = 800$. The expected Kolmogorov scaling of the power spectrum is seen in the inertial regime at earlier times. In the inviscid case, the numerical viscosity has a minor damping effect on the power spectrum at late times and high wavenumbers (short lengthscales). A greater damping effect is seen in the viscous case, as well as a `ringing' at high wavenumbers due to coupled action of fluid element discretization and local viscosity: these wavenumbers correspond to lengthscales of a few, or even a single, cell(s).}
    \label{fig:KHI_Kolmo_Compare}
\end{figure}
\Cref{fig:KHI_Kolmo_Compare} shows the power spectra for the kinetic energy in our KHI simulations. Two comparisons are made: one between early ($t=12.0$, blue curve) and late ($t=30.0$, orange curve) times; and one between an inviscid (top panel) and viscous (bottom panel) fluid. In both cases, the system loses energy over time. For the inviscid case, this is due to numerical dissipation. For the viscous case, there is the additional effect of viscous dissipation, which causes the steeper drop-off for the orange vs. the blue curve. The expected Kolomogorov scaling for the inertial range is plotted and matches well with the data for all but the late-time viscous case, where dissipation has more efficiently moved energy to the shorter lengthscales, giving a steeper dependence on wavenumber. Finally, the `ringing' effect seen for the highest wavenumbers in the late-time, viscous case is, we believe, a numerical artefact. Dissipative behaviour in the MISCE model is captured using a complex mixture of spatial gradients of the primitive variables, generally calculated using simple central differencing rather than, for example, a WENO scheme which is designed to be non-oscillatory. For the  highest wavenumbers here, corresponding to a few or even a single cell(s), these derivatives may be causing small-scale oscillations in fluid variables that have no qualitative significance.

\subsection{Code Performance}
\label{subsec:code_performance}

There are a few key metrics of code performance we must now consider. Firstly, how the runtime of simulations scales with resolution. Secondly, the convergence of the simulation output, which is assessed in two ways: the self-convergence of the MISCE results to a very high resolution simulation output; the asymptotic approach of the MISCE results to either leading-order or ideal fluid behaviour as the non-ideal coefficients approach zero. Finally, we present a comparison of runtimes, showing the significant speed-up achieved by the MISCE model.

\subsubsection{Scaling and Convergence}
\label{sec:scaling_convergence}

\begin{figure}
    \centering
    \includegraphics[scale=0.58]{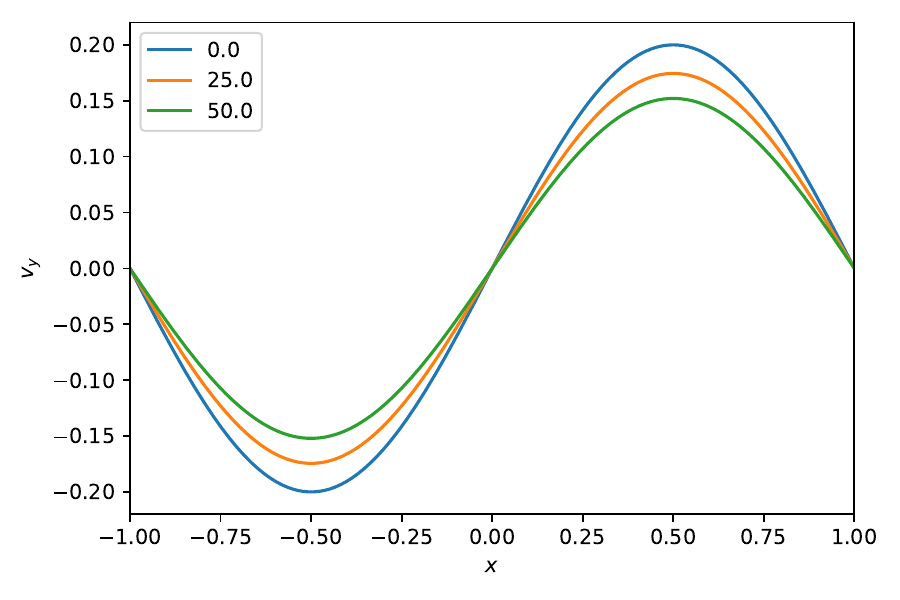}
    \caption{The ``SineWave'' initial data (and its evolution with the MISCE model) used to assess numerical convergence with resolution. Shear viscosity causes the flattening of features in the $y$-direction velocity across the $x$-domain here. This simulation was performed using $3200$ cells in one dimension up to a code time of $50.0$.}
    \label{fig:sinwave}
\end{figure}

By evolving smooth ``SineWave'' initial data~(\cref{fig:sinwave}), we are able to assess the convergence of our MIS and MISCE implementations with resolution. Considering the error due to finite resolution, we define it as the difference between ``true'' solution (the one obtained at infinite resolution) and the finite-resolution solutions our code actually produces: $\mathcal{E} = Q_{true} - Q_{num}$. 

Then we make the usual assumption that this error follows a power law scaling in the grid-size: $\mathcal{E} \propto \Delta x^n$. Different approaches exist for extracting the value of $n$ in this expression, and we choose here to use self-convergence, where each resolution's solution is compared to its neighbours to produce a set of convergence powers at different resolutions. We do this because different components of our numerical scheme (the time-integrator, cell-interface reconstruction method etcetera) each have individual expected convergence rates that blend together to give an overall convergence. This means that different components can dominate the error at different resolutions, and we are able to assess the transition between them using this approach. 

We show in~\cref{tab:selfconvergence} a summary of convergence orders for different models and resolutions. In summary, we see a transition from high-order convergence at low resolutions to lower-order convergence at higher resolutions. For both models, the error at low resolutions is dominated by the time integrator and reconstruction algorithm, which are high-order schemes and hence their error converges away quickly. 

At high resolutions, we see a drop in the overall convergence order. For the MISCE model, this is because there are many spatial derivatives of the primitive variables in the complex source terms, which are evaluated using second-order central differencing. Increasing the order of this central differencing does increase the convergence rate, but makes negligible difference to quantitative results. For the MIS model, we require temporal derivatives of the primitive variables. These are evaluated using backwards-differencing on the primitives' values at the current and previous timestep. This introduces a first-order error due to these lagging updates that does not converge away with resolution, and hence appears as the dominant error at high resolutions.

\begin{figure}
    \centering
    \includegraphics[scale=0.6]{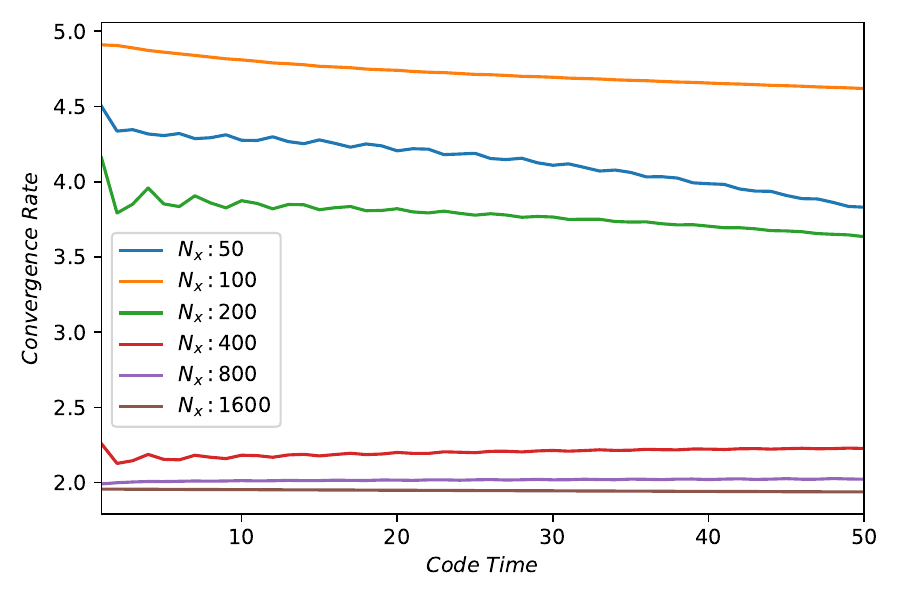}
    \caption{The self-convergence of the MISCE model  across time for a range of resolutions. The simulation data can be seen in~\cref{fig:sinwave}, although this convergence test is carried out on the number density, $n$. For lower resolutions, the convergence order is between fourth and fifth due to the use of an RK4 time-integrator and a WENO5 reconstruction scheme. At higher resolutions a transition to second-order convergence is seen due to the presence of first-order central differencing used for spatial derivatives in the MISCE source terms.}
    \label{fig:Coarse_Convergence_SimpleCD_n_MISCE_Shear_HHO}
\end{figure}
\begin{figure}
    \centering
    \includegraphics[scale=0.65]{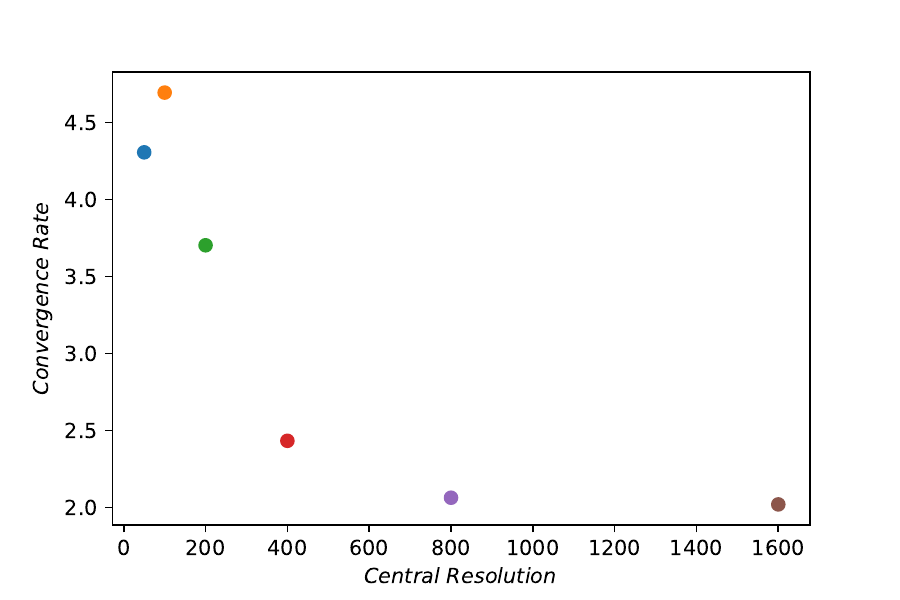}
    \caption{The self-convergence rate of the MISCE model for different resolutions. For lower resolutions, the convergence order is between fourth and fifth due to the use of an RK4 time-integrator and a WENO5 reconstruction scheme. At higher resolutions a transition to second-order convergence is seen due to the presence of first-order central differencing used for spatial derivatives in the MISCE source terms.}
    \label{fig:ConvergenceVsResolution_SimpleCD_n_MISCE_Shear_HHO}
\end{figure}
\begin{table}
    \centering
    \begin{tabular}{c c c c c}
    & \multicolumn{3}
    {c}{\textbf{Self-Convergence}} & \\
         Model & Integrator & Reconstruction & Resolution & Order  \\ \hline
          MISCE & RK2 & WENO3 & 100 & 2.45 \\ \hline
          MISCE & RK4 & WENO5 & 50 & 4.5 \\      MISCE & RK4 & WENO5 & 100 & 4.7 \\
          MISCE & RK4 & WENO5 & 200 & 3.9 \\
          MISCE & RK4 & WENO5 & 400 & 2.6 \\      MISCE & RK4 & WENO5 & 800 & 2.1 \\    
          MISCE & RK4 & WENO5 & 1600 & 2.0 \\ \hline     
          MIS & SSP2 & WENO5 & 400 & 4.0 \\
          MIS & SSP2 & WENO5 & 800 & 5.7 \\
          MIS & SSP2 & WENO5 & 1600 & 1.0 \\
    \end{tabular}
    \caption{The self-convergence of a smooth sin-wave evolution using different models of non-ideal hydrodynamics and different numerical schemes. The expected orders of convergence are seen. At very high resolution, the first-order  central differencing used in the MISCE model source's spatial derivatives causes the convergence order to drop to $2^{nd}$. For the MIS model, we use lagged-updates to calculate the required time derivatives. This similarly caps the order of convergence at first when very high resolutions are used and error from other components of the numerical scheme are tiny.}
    \label{tab:selfconvergence}
\end{table}

\subsubsection{Model Comparison}

\begin{table}
    \centering
    \begin{tabular}{c c|c c|c}
    & & \multicolumn{2}  
    {c}{\textbf{Average Runtime}} &  \\
         Model & Integrator & Resolution & Endtime & Runtime (Speed-up)  \\ \hline
          MIS & SSP2 & 200x400 & 6.25 & 1h22m \\ 
          MISCE & RK2 & 200x400 & 6.25 & 6m ($\times$14) \\ \hline 
          MIS & SSP2 & 400x800 & 6.25 & 3h22m \\ 
          MISCE & RK2 & 400x800 & 6.25 & 29m ($\times$7) \\ \hline 
          MIS & SSP2 & 800x1600 & 6.25 & 26h10m \\
          MISCE & RK2 & 800x1600 & 6.25 & 3h7m ($\times$8.4) \\ \hline
          MIS & RK2 & 800x1600 & 18.0 & 15h45m \\
          MISCE & RK2 & 800x1600 & 18.0 & 9h8m ($\times$1.7) \\ \hline
          MIS & SSP2 & 800x1600 & 3.75 & 22h3m \\
          MISCE & RK2 & 800x1600 & 3.75 & 1h54m ($\times$11.6) \\
    \end{tabular}
    \caption{A comparison of computational time required for different hydrodynamic models and time-integrators. These results are for Kelvin-Helmholtz instability simulations using 40 CPU nodes and MPI memory management on the Iridis5 supercomputer. The MISCE model gives about an order of magnitude speed-up compared to the MIS model (when evolved with explicit methods instead of implicit ones). RK2 refers to an operator-split, $2^{nd}$-order Runge-Kutta scheme and SSP2(222) refers to a $2^{nd}$-order implicit-explicit scheme.}
    \label{tab:KHI_runtimes_model_compare}
\end{table}
We show in~\cref{tab:KHI_runtimes_model_compare} a comparison of runtimes between the MIS and MISCE models for the KHI. We primarily present results comparing the MIS model evolved with the SSP2(222) IMEX time-integrator~\citep{pareschi_implicitexplicit_2005} and the MISCE model evolved with an operator-split RK2 time-integrator. Whilst this comparison may seem `unfair' at first, due to the costly nature of IMEX schemes compared to explicit ones, it is justified. Whilst for much of a merger simulation the neutron star fluid may be accurately treated as ideal or near-ideal, when dissipation does become significant its parameter space will certainly extend into the region where the MIS model becomes stiff and IMEX schemes are needed to evolve it stably.

In this case, a significant speed-up of about an order of magnitude is achieved using the MISCE model. When the two models are compared using explicit time-integrators for both, a speed-up of nearly a factor of $2$ occurs, owing to the reduced system size.

\section{Conclusions}
\label{sec:conclusions}

We have presented a dissipative extension to the relativistic, ideal hydrodynamic equations often used in astrophysical simulations. Motivated by the relaxation form of the MIS sources for the dissipative variables,  new source terms are derived by  writing the dissipative variables as a series expansion in deviations from their equilibrium, relativistic-Navier-Stokes values. The series is paramterized by the dissipation strength and timescale coefficients and its terms are found using an order-by-order comparison of the MIS equations of motion. This leads to a rapidly convergent series in the case of fast-acting, weak dissipation, which we term the MISCE formulation.

This new system is numerically non-stiff in the exact limit where the commonly-used MIS equations of motion are stiff, and vice versa. Because much of the matter in a neutron star may be treated as a near-ideal fluid, the MISCE equations of motion are often able to be evolved explicitly, giving accurate results with execution times that are about an order of magnitude reduced. Even when both models are evolved with the same, explicit integrators, the MISCE formulation is nearly twice as fast, owing to its reduced system size. It also converges to the  Euler equations in the zero-dissipation limit, allowing for the natural evolution of a fluid which is mostly ideal with some areas of non-equilibrium behaviour.

Within its domain of validity, we have demonstrated it to produce highly similar results to the MIS formulation for a range of initial data. It is able to capture dissipative effects near discontinuous data without the onset of Gibbs oscillations, and shows little error growth (compared to MIS results) for smooth solutions over dissipation strengths and timescales spanning many orders of magnitude. For more complex simulations of Kelvin-Helmholtz instabilities, the expected scaling laws are reproduced for the kinetic power spectrum across the inertial range of wavenumbers. 
	
The realm of stability for our new model is considered in \cref{sec:stability_analyis}, and is dependent both upon the dissipation coefficients (in particular the ratio of strength to timescale) and the simulation's spacetime resolution, with a sharper dependence on the spatial resolution. The presence of many mixed-order derivatives in the source terms can lead to instabilities when spatial resolutions are pushed very high, though this effect may be mitigated somewhat by using better numerical-derivative approximations (than simple finite-differences) such as slope-limiting ones. In~\cref{sec:scaling_convergence} we presented results showing the expected convergence for the fourth-order Runge-Kutta and fifth-order WENO schemes we use. One caveat is that at high resolutions, the MISCE formulation, which makes use of second-order central-differencing of the primitive variables, starts converging at second-order in the grid-spacing. Similarly, the MIS formulation starts converging at first-order for high resolutions, when the dominant error contribution is the first-order time derivatives calculated using lagged updates. 

In \cref{sec:MISCE_Derivation}, to simplify the form of the matrices in the source that we present, we made the assumption that terms of $\mathcal{O}(v^2)$ and higher were negligible, and hence that the Lorentz factor, $W$, could be set to unity. Whilst for simulations we use the entire, non-simplified expressions that we derived using computer algebra, the differences this made to results were small, and were generally eclipsed by resolution effects. However, the differences may be more significant for fluid velocities approaching the speed of light, such as in the final orbits of a binary neutron star pre-merger, or for the significantly out-of-equilibrium matter created in the merger itself.

Although all simulations have been performed in the special relativistic limit, the techniques we have used are not limited to this alone. A general relativistic extension to MISCE is (in principle, at least) straightforward and already underway. In addition, we have developed an adaptive code prototype that evolves different dissipative fluid formulations in different physical regimes (e.g. MIS and MISCE) to minimize computational work and maximize accuracy and stability. Together, this should allow for more efficient, dissipative simulations of neutron star mergers and accretion onto compact objects.

\section*{}

The authors acknowledge the use of the IRIDIS High Performance Computing Facility, and associated support services at the University of Southampton, in the completion of this work. Open source software used includes SymPy \citep{meurer_sympy_2017}, Matplotlib \citep{hunter_matplotlib_2007} and CMINPACK (\href{http://devernay.free.fr/hacks/cminpack/}{Devernay 2017}). 

\appendix

\section{Rapid Evolution of Reduced Initial Data}
\label{app:bound_layers}

The source terms in MIS models drive the dissipative variables towards their equilibrium values on timescales $\tau$. We therefore expect that, when our initial data (or otherwise) puts us significantly out-of-equilibrium, at times $t \lesssim \tau$, there will be a systematic error in the dissipative variables that decays roughly as $e^{-t/\tau}$. 

However, a CE-expanded model does not possess this type of source term nor indeed any explicit dissipative variables at all. Instead, the primitive variables and their derivatives are used to produce dissipative effects. We therefore expect that we will need to make modifications to the primitive variables' initial values to reflect their out-of-equilibrium status in lieu of having terms that explicitly define our out-of-equilibrium state.

\title{Simple Heat Model}
\label{subsec:BL_ToyQ}

Let us demonstrate the effect of not making appropriate adjustments to the primitive variables to reflect their out-of-equilibrium state. We take the simple heat model presented earlier in~\cref{eq:toy_heat}, in one dimension:
\begin{subequations}
    \begin{align}
        \partial_t T + \partial_x q &= 0,     \label{eq:toy_heat_BL_a} \\ 
        \partial_t q &= -\frac{1}{\tau_q} (\kappa \partial_x T + q);     \label{eq:toy_heat_BL_b}
    \end{align}
    \label{eq:toy_heat_BL}
\end{subequations}
and its CE form, \cref{eq:toy_heat_CE_simple},
\begin{equation}
    \partial_t T = \kappa \left[ \partial^{(2)}_x T - \kappa \tau_q \partial^{(4)}_x T \right].
    \label{eq:toy_heat_CE_BL}
\end{equation}
By introducing a fast time variable $\mathcal{T} = t / \tau_q$ on the scale of the relaxation rate, we can perform a matched asymptotic expansion valid even at small times. This transforms~\cref{eq:toy_heat_BL} into 
\begin{subequations}
    \label{eq:toy_heat_eps}
    \begin{align}
        \label{eq:toy_heat_eps_a}
        \partial_\mathcal{T} T + \tau_q \partial_x q &= 0, \\
        \label{eq:toy_heat_eps_b}
        \partial_\mathcal{T} q &= -q - \kappa \partial_x T.
    \end{align}
\end{subequations}
From~\cref{eq:toy_heat_eps}, the power series expansion now gives that the temperature $T$ is independent of $\mathcal{T}$ to leading order and 
\begin{equation}
    \partial_\mathcal{T} q_0 = -q_0 
\end{equation}
which can be integrated directly to give
\begin{equation}
    q_0 = C_0 e^{-\mathcal{T}} 
    \label{eq:q_0_decay}
\end{equation}
where $C_0$ is a constant of integration. We immediately see that this exponential behaviour in fast time, $\mathcal{T}$, cannot be captured by a power series expansion in the original time, $t$.

Noting that $C_0 = q(t = 0) + \mathcal{O}(\tau_q)$, we relabel $C_0$ as $\Delta q_0$ because it represents an initial offset of the heat-flux at $\mathcal{T} \to 0^+$. To compare this early-time behaviour between the two models (\cref{eq:toy_heat_BL} and \cref{eq:toy_heat_CE_BL}), we can Taylor-expand each in terms of $\mathcal{T}$ to an arbitrary (small) time $\mathcal{T} = 1$ about $\mathcal{T} = 0$. This is equivalent to considering the evolution up to time $t=\tau_q$. From the relaxation model~\cref{eq:toy_heat_BL} we have
\begin{subequations}
    \begin{align}
        T(\mathcal{T}=1) &\simeq T(\mathcal{T}=0) +  \left(\frac{\partial T_0}{\partial \mathcal{T}} + \tau_q \frac{\partial T_1}{\partial \mathcal{T}} + ... \right) (\mathcal{T}=0) + ... \\
        &= T(\mathcal{T}=0) - \tau_q \frac{\partial \Delta q_0}{\partial x} + \mathcal{O}(\tau_q^2)
    \end{align}
\end{subequations}
whilst from the CE model~\cref{eq:toy_heat_CE_BL} we have
\begin{subequations}
    \begin{align}
        T(\mathcal{T}=1) &\simeq T(\mathcal{T}=0) +  \left(\frac{\partial T_0}{\partial \mathcal{T}} + \tau_q \frac{\partial T_1}{\partial \mathcal{T}} + ... \right) (\mathcal{T}=0) + ... \\
        &= T(\mathcal{T}=0) + \mathcal{O}(\tau_q^2).
    \end{align}
\end{subequations}
Comparing the two, we see that we can match the two temperatures at small times by making an initial-data adjustment given by
\begin{equation}
  \label{eq:toy_heat_BL_correction}
    T(t=0,x) \to T(t=0,x) - \tau_q \partial_x \Delta q_0.
\end{equation}
This accounts for the fast-relaxation behaviour and removes the exponentially-decaying, leading-order systematic error in the solution. In words, we are taking into account the heat flux ($q$) that would have produced our initial, out-of-equilibrium temperature ($T$) distribution. Otherwise, our reduced system does not have access to this knowledge and will not equilibrate accordingly.

\begin{figure*}
    \centering
    \includegraphics[scale=0.65]{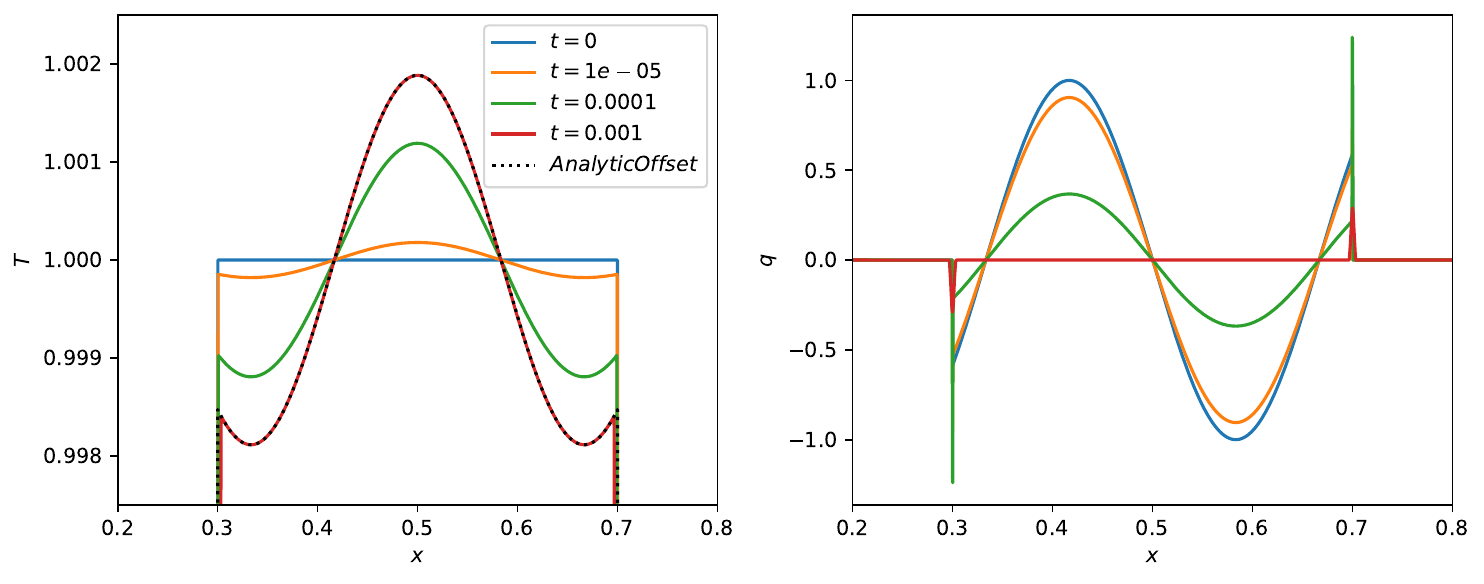}
    \caption{The evolution of the temperature and heat flux for the initial data described in~\cref{app:bound_layers} using the relaxation model given by~\cref{eq:toy_heat_BL}. The initial heat flux means the data is initially out-of-equilibrium. The non-ideal parameters are $\kappa = \num{1e-3}$ and $\tau_q = \num{1e-4}$, so the system relaxes to equilibrium on the timescale shown here, as seen by the heat flux relaxing to nearly zero. The analytic result for the appropriate adjustment to the initial data, derived in~\cref{app:bound_layers}, is also plotted in the left panel (dotted) and shows excellent agreement with the numerical evolution result.}
    \label{fig:ToyQ_BL_Comp}
\end{figure*}

This can be seen in \cref{fig:ToyQ_BL_Comp}, where a one-dimensional `top-hat' temperature profile evolves up to $t = \num{1e-3}$, using non-ideal parameter values $\kappa = \num{1e-3}$ and $\tau_q = \num{1e-4}$. Here, we are not interested in the usual, long-term evolution where heat would slowly diffuse outwards and the temperature profile would adopt a decaying Gaussian shape. Instead, we are interested in the very short-term evolution due to the inclusion of an initial heat flux $\Delta q_0 = \sin(6 \pi x)$ into the MIS-derived heat model given by \cref{eq:toy_heat_BL}. 

In the left panel the temperature of the MIS-derived relaxation model is shown with solid lines, whilst the initial temperature with the offset computed in~\cref{eq:toy_heat_BL_correction} is shown with dotted lines. Excellent agreement is seen, indicating that this offset would work when applied to a reduced order model such as the MISCE approach. The right panel shows the heat flux of the MIS-derived relaxation model, showing that the system has relaxed to equilibrium, illustrating that this applied offset has the appropriate magnitude.

In summary, injecting an initial heat-flux into a relaxation-type system leads to an exponentially fast adjustment of the corresponding conjugate primitive variable: the temperature, in this case. We are able to derive an analytic expression for this adjustment behaviour that depends on the spatial gradient of the injection and the non-ideal parameter controlling its timescale, $\tau_q$ in this case. Even in the reduced system found using the CE-expansion, we are able to adjust the sole remaining variable (the temperature) to capture the offset that is quickly arrived at by the original relaxation system. 

The same effect would be observed when using the full MIS model of non-ideal hydrodynamics, where an initial bulk or shear viscosity would lead to an exponentially-fast adjustment of the velocity, albeit likely small in magnitude. If one uses our MISCE model for capturing far out-of-equilibrium dissipation, the initial conditions of the non-ideal variables (viscosity, heat-flux) can and should still be taken account of by adjustment of their conjugate primitive variables (such as velocity and temperature).

\section{Stability Analysis}
\label{sec:stability_analyis}

It is important to consider the numerical stability of the CE systems introduced here. Usually, conservation laws are evolved for hydrodynamic simulations of ideal fluids in special relativity of the form
\begin{equation}
    \partial_t q + \partial_x f = 0,
\end{equation}
where we choose to write it in one spatial dimension for simplicity. The Courant–Friedrichs–Lewy (CFL) condition sets a stability criterion for these strongly hyperbolic systems given by
\begin{equation}
    C := \abs{\frac{\partial f}{\partial q}} \frac{\Delta t}{\Delta x} \leq C_{max}
\end{equation}
where $\Delta t$ is the timestep and $\Delta x$ is the spatial resolution. $C$ is the Courant number and $C_{max}$ is a constant that determines its maximum stable value and depends on the particular numerical scheme employed. Typically, $C_{max} = \mathcal{O}(1)$ for explicit schemes.

\begin{figure}
    \centering
    \includegraphics[scale=0.525]{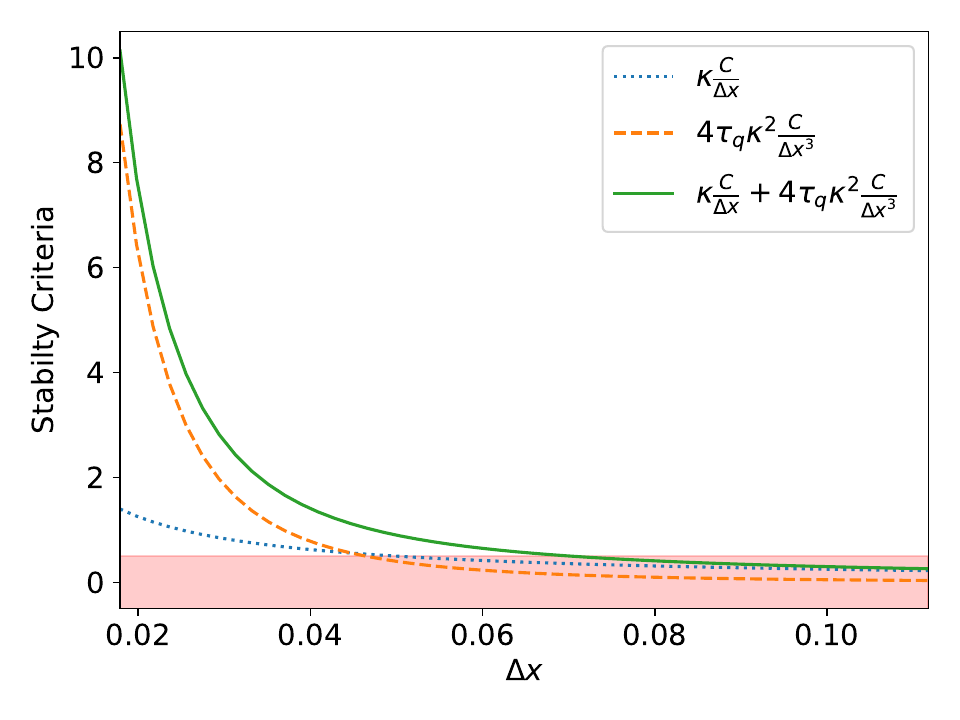}
    \caption{The stability criteria of~\cref{eq:toyq_CE_overall_vN_stability} are plotted separately (dashed, dotted) and summed (solid). The red, shared region shows where the simple heat flux model given by \cref{eq:toy_heat_CE_for_stability} should be stable, according to the standard von Neumann analysis technique using a Fourier series of errors. The heat dissipation parameter values here are $\tau_q = 0.01$, $\kappa = 0.05$ and the CFL factor is $C = 0.5$ hence the crossover between the two stability criteria occurs at $\Delta x \simeq 0.045$.}    \label{fig:toy_heat_CE_stability}
\end{figure}

For linear systems involving higher-order spatial derivatives, such as the heat equation given by
\begin{equation}
  \label{eq:heat}
    \partial_t T = \kappa \partial^{(2)}_x T,
\end{equation}
a von Neumann stability analysis is usually performed, where the numerical errors are decomposed into a Fourier series. For a forward-time, centre-spaced (explicit) numerical scheme this analysis finds that 
\begin{equation}
    \kappa \frac{\Delta t}{\Delta x^2} \leq \frac{1}{2}
\end{equation}
is the condition necessary for stability. For the CE heat system given by~\cref{eq:toy_heat_CE_simple},
\begin{equation}
  \label{eq:toy_heat_CE_for_stability}
    \partial_t T = \kappa \partial^{(2)}_x T - \kappa^2 \tau_q \partial^{(4)}_x T,
\end{equation}
the conditions for stability are more complex. See~\cite{bevilacqua_significance_2011} for a treatment of higher-order differential terms relevant to our work here. 
The analysis gives us the inequality for stability of
\begin{equation}
    \label{eq:toyq_CE_overall_vN_stability}
    \kappa \frac{\Delta t}{\Delta x^2} + 4 \tau_q \kappa^2 \frac{\Delta t}{\Delta x^4} \leq \frac{1}{2}.
\end{equation}
This clearly yields the previous heat-equation limit when the first term is dominant. In the limit where $\Delta x \to 0$ and the second term becomes dominant we instead obtain the condition $\Delta t \leq \frac{1}{8} \frac{\Delta x^4}{\tau_q \kappa^2}$ which is the stricter condition of the two in this limit. We expect a cross-over of stability between the two criteria when $\Delta x^2 = 4 \tau_q \kappa$. At this point, the overall stability condition given by~\cref{eq:toyq_CE_overall_vN_stability} above yields $\Delta t \leq \tau_q$. See~\cref{fig:toy_heat_CE_stability} for a visualisation of these stability criteria.

The stability criteria for the full, non-linear MISCE model we have developed will be more complex still, given the presence of many mixed derivatives. Hence, we primarily investigate its stable parameter space empirically. However, we can first glean some insight analytically, although the usual von Neumann stability analysis is not applicable to the non-linear terms and we therefore consider the linear terms only here.

We make the ansatz that the solution can be written as $q^n_l = q^n \exp(il\alpha \Delta x)$ where $n$ and $l$ index the time-step and a grid-point, respectively, and $\alpha$ is a spatial frequency present in the data. Using central finite differencing, the MISCE sources will produce a solution growth rate per step, $q$, with the following form
\begin{align*}
    q = 1& - \xi\frac{A \Delta t}{\Delta x^2}\sin^2{\left(\frac{\theta}{2}\right)} + \xi \tau\frac{B \Delta t}{\Delta x^3} \left(\sin{(2 \theta)} - 2\sin{(\theta})\right) \allowbreak \\ &+ \xi^2 \tau\frac{C \Delta t}{\Delta x^4}\sin^4{\left(\frac{\theta}{2}\right)}
\end{align*}
where $\theta = \alpha \Delta x/2$, $\xi \equiv \{\zeta, \kappa, \eta \}$ and $A$, $B$ and $C$ are functions of the primitive variables. We anticipate a crossover between the various stability criteria as resolution varies. 

Firstly, we note that the validity of our expansion only applies when $\tau \ll 1$ and indeed we find that our simulations are unstable when $\tau \gtrsim \num{e-2}$. In~\cite{hiscock_stability_1983}, conditions are provided for the stability (and causality) of MIS theory. There, the $\beta$ coefficients given in~\cref{eq:betas}, which represent the ratio of dissipation strengths ($\xi \equiv \{\zeta,\kappa,\eta\}$) to timescales ($\tau \equiv \{\tau_\Pi,\tau_q,\tau_\pi\}$), are used to determine the stability of the theory. Unsurprisingly then, we find the same is true here: a lower bound from causality appears on the ratio $\tau/\xi$. 

In particular we find that for shocktube tests, $\tau_q/\kappa = T\beta_1 \gtrsim 0.5$ as well as $\tau_\Pi/\zeta = \beta_0 \gtrsim 0.1$ grants stability. Similarly, for KHI tests, $\tau_\pi/\eta = \beta_2 \gtrsim 0.2$ gives stability. These ratio conditions coupled with the small-$\tau$ requirement work together to create a stability region bounded at either end, at least when next-to-leading order terms $\propto \tau$ are included in the MISCE source.

For a Boltzmann gas, the $\beta$ coefficients have thermodynamic forms that we can calculate analytically. We expect them to usually be $\sim \mathcal{O}(1)$. Indeed, we have implemented these thermodynamic forms such that the timescales used are dynamically adjusted during the simulation -- little difference is made to using preset values.

\section{Approximating Time Derivatives}
\label{app:dwdt_NLO}

In~\cref{sec:MISCE_Derivation} we showed that by making use of the fundamental conservation-law equation
\begin{equation}
    \tag{\ref{eq:cons law ZO}}
        \partial_t {\bm{q}}(\bm{w}) + \partial_i \bm{f}^{i}(\bm{w}) = \bm{0}
\end{equation}
and the simple chain-rule for derivatives, we are able to arrive at an approximation to time derivatives of the primitive variables, ${\bm w}$, containing only spatial derivatives of the fluxes (or primitive variables). 

\begin{equation}
    \frac{\partial \bm{w}}{\partial t} = \frac{\partial \bm{w}}{\partial \bm{q}} \frac{\partial \bm{q}}{\partial t} = -\left(\frac{\partial \bm{q}}{\partial \bm{w}}\right)^{-1} \partial_i\bm{f}^{i} = -\bm{A}\bm{B}
\end{equation}
where 
\begin{equation}
    \label{eq:A&B_simplest}
    A = \left(\frac{\partial \bm{q}}{\partial \bm{w}}\right)^{-1}
    , ~ B = \partial_i\bm{f}^{i}.
\end{equation}

In the case of BDNK models of dissipative fluids, and of our MISCE model presented here (to first-order), we may write both the conservative and flux vectors in an expanded form that separates (first-order) derivatives in the primitive quantities:

\begin{align}
        &\partial_t \left[ \bm{q}_{(0)}(\bm{w}) + \epsilon \bm{q}_{(1)}(\bm{w}, \partial_t\bm{w}, \partial_i\bm{w}) \right] \\ &+ \partial_i \left[ \bm{f}_{(0)}^{i}(\bm{w}) + \epsilon \bm{f}_{(1)}^{i}(\bm{w}, \partial_t\bm{w}, \partial_i\bm{w}) \right] = 0 \notag
\end{align}
where $\epsilon$ parametrizes the size of dissipation in the fluid model. Under the assumption that dissipation is small compared to the bulk behaviour of the fluid, $\epsilon$ is small. This is the regime when using the MISCE makes sense, anyway. 

Now we cannot only consider the contribution of the fluid variables themselves to the time-derivative of the state vector ($\partial_t \bm{w}$), but also the contribution of the temporal and spatial derivatives ($\dot{\bm{w}}$, $\bm{w}'$). We first rewrite $\bm A$ and $\bm B$ as
\begin{equation}
    \label{eq:A&B_simplest_rewrite}
    A = \left[ \partial_w {\bm q}_0 + \epsilon \partial_{\bm w} {\bm q}_1 \right]^{-1}
    , ~ B = \left[ \partial_i {\bm f}^i_0 + \epsilon \partial_i {\bm f}^i_1 \right].
\end{equation}
After much manipulation, and using the assertion that $\epsilon$ is indeed small to expand a sum of matrices to leading order, we arrive at
\begin{subequations}
  \begin{align}
    \label{eq:A&B_expanded}
    A = \left[ \left({\bm I} - \epsilon \left(\frac{\partial {\bm q}_0}{\partial w}\right)^{-1} \frac{\partial {\bm q}_1}{\partial {\bm w}} \right) \left(\frac{\partial {\bm q}_0}{\partial {\bm w}}\right)^{-1} \right], \\
    B = \left[ \frac{\partial {\bm f}_0^i}{\partial {\bm w}} {\bm {\bm w}}' + \epsilon \left( \frac{\partial {\bm f}^i_1}{\partial {\bm w}}{\bm w}' + \frac{\partial {\bm f}^i_1}{\partial \dot{{\bm w}}}\dot{{\bm w}}' + \frac{\partial {\bm {\bm f}}^i_1}{\partial {\bm w}'}{\bm w}'' \right) \right].
  \end{align}
\end{subequations}
Note that we can choose to use the form of $A$ in \cref{eq:A&B_simplest} and invert the sum of matrices, rather than using the approximate small-$\epsilon$ trick that leads to $A$ in \cref{eq:A&B_expanded}. Similarly, we can choose the expression for $B$ from \cref{eq:A&B_simplest} which makes use of the fluxes themselves directly, or we may use its form in \cref{eq:A&B_expanded} which requires evaluation of second-order spatial differences of the primitives.




\bibliographystyle{mnras}
\bibliography{DEIFY} 

\bsp	
\label{lastpage}
\end{document}